\documentclass[aps,prd,twocolumn,nofootinbib,superscriptaddress,10pt,floatfix]{revtex4-2}
\usepackage{amsmath,amssymb,mathtools,bm}
\usepackage{graphicx}
\usepackage{booktabs,multirow,array}
\usepackage[dvipsnames]{xcolor}
\usepackage[colorlinks=true,linkcolor=MidnightBlue,citecolor=BrickRed,urlcolor=RoyalBlue]{hyperref}
\usepackage{microtype}
\usepackage{placeins}
\newcommand{\Mp}{M_{\rm P}}

\newcommand{\Tr}{\operatorname{Tr}}
\newcommand{\BR}{\operatorname{BR}}
\newcommand{\diag}{\operatorname{diag}}
\newcommand{\Eq}[1]{Eq.~\eqref{#1}}
\newcommand{\Eqs}[2]{Eqs.~\eqref{#1} and~\eqref{#2}}
\newcommand{\Fig}[1]{Fig.~\ref{#1}}
\newcommand{\Sec}[1]{Sec.~\ref{#1}}
\newcommand{\Tab}[1]{Table~\ref{#1}}
\begin{document}
\title{Inverse-seesaw and inflationary scale relations in a supersymmetric 331 model from an \texorpdfstring{$SU(6)$}{SU(6)} GUT model}
\author{Imtiaz Khan}
\email{ikhanphys1993@gmail.com}
\affiliation{Department of Physics, Zhejiang Normal University, Jinhua, Zhejiang 321004, China}
\affiliation{Research Center of Astrophysics and Cosmology, Khazar University, Baku, AZ1096, 41 Mehseti Street, Azerbaijan}

\author{Tianjun Li}
\email{tli@itp.ac.cn}
\affiliation{School of Physics, Henan Normal University, Xinxiang 453007, P. R. China}

\date{\today}

\begin{abstract}
We propose a $N=1$ supersymmetric $SU(6)$ theory with an intermediate $(SU(3)_C\times SU(3)_L\times U(1)_X)$ phase, in which common fields and operators relate the neutrinos, gauge mediation, unification, and symmetry-breaking sectors. The antisymmetric Dirac matrix of the inverse-seesaw sector
has rank two, yielding one vanishing light-neutrino mass at leading order. A higher-dimensional symmetry-breaking correction lifts this zero mode through its projection
onto the left and right null vectors of the Dirac matrix, leading to a quartic suppression in the ratio of the intermediate and ultraviolet scales. Complete unified
messenger multiplets preserve the relative one-loop gauge coupling evolution while modifying the unified coupling. When the same supersymmetry-breaking spurion
generates the singlet lepton-number-violating parameter, an analytical relation emerges between the messenger scale, the lepton-number mediator scale, and the unified
coupling. The two-loop gauge evolution remains perturbative for the assumed heavy-threshold spectrum. In the inflationary sector, the scalar-amplitude normalization
determines the coupling governing both the inflationary potential and the radial scalar mass in the broken vacuum. The physical scalar directions transverse to
the inflaton are stable within the neutral supergravity subspace, while the combined inflationary and gauge-mediated conditions imply a significant sensitivity
of the symmetry-breaking vacuum. An examination of the allowed operators shows that additional ultraviolet conditions are required to obtain the observed active-neutrino
mixing, suppress competing lepton-number-violating terms, and complete the heavy-field spectrum. Our analysis
establishes the analytical scale relations among the neutrino, messenger, unification, and inflationary sectors and identifies the conditions necessary for their
simultaneous realization.
\end{abstract}
\maketitle

\section{Introduction}\label{sec:intro}
In supersymmetric grand unified theories, messenger masses, intermediate symmetry-breaking scales, lepton-number violation, heavy thresholds, and inflationary couplings are generally independent parameters.  Gauge-coupling unification is sensitive to the superpartner spectrum and to heavy thresholds~\cite{DimopoulosGeorgi1981,Sakai1981,Amaldi1991,LangackerLuo1991}, and precision matching further specifies this threshold dependence~\cite{LangackerPolonsky1993,RossRoberts1992,BaggerMatchevPierce1995,Pierce1997}.  Relations among these parameters may arise when the same fields or operators occur in otherwise separate sectors.  A supersymmetric $331$ model embedded in an $SU(6)$ grand unified theory provides a setting in which to examine such relations.  Its unified multiplets determine the $331$ embedding and contain the fields entering the inverse seesaw, gauge mediation, and the symmetry-breaking direction.

Models based on the $331$ gauge structure have been studied with different anomaly-free family assignments, scalar sectors, charge embeddings, and collider signatures~\cite{Singer1980,PisanoPleitez1992,Frampton1992,Montero1993,FootHernandez1993,PleitezTonasse1993,FootLongTran1994,Long1996,Okada2016,CorianoMelle2024}.  An $SU(6)$ embedding determines the Abelian normalization and accommodates each chiral family in a small set of unified representations~\cite{Deppisch2016}.  In particular, the $\mathbf{15}+2\overline{\mathbf6}$ family content was used to obtain a rank-two inverse seesaw in a nonsupersymmetric $331$ model from the $SU(6)$ GUT model~\cite{Le2020}.  We retain this group-theory structure and include supersymmetry, complete gauge-messenger multiplets, and a $331$-breaking pair that also parametrizes the inflationary trajectory.  We study whether these relations remain valid for the physical light states and the complete symmetry-allowed field content.

The inverse-seesaw mechanism allows neutral fermions at the TeV scale, with lepton-number violation controlled by a small Majorana matrix~\cite{Mohapatra1986,DeppischValle2005}.  Antisymmetric Dirac matrices occur in $331$ models~\cite{Dias2012}, and their flavor and nonunitarity effects have been examined in supersymmetric and low-scale seesaw scenarios~\cite{DevMohapatra2010,ParkWang2011,IlakovacPilaftsis2009,IlakovacPilaftsisPopov2013,Abada2014,Boucenna2014,Vicente2015,KriewaldTeixeira2025}.  Higher-dimensional operators may perturb antisymmetric $SU(6)$ neutrino textures~\cite{ChackoDevMohapatraThapa2020}.  For a conditional inverse-seesaw texture, we express the first nonzero Takagi mass in terms of the left and right null vectors of its rank-two Dirac matrix.  This expression specifies the flavor projection responsible for rank breaking and shows that the lightest mass is quadratic in the leading Dirac-matrix correction.

The Higgs sector of supersymmetric $SU(6)$ has also been considered in pseudo-Goldstone and missing-VEV constructions, where doublet--triplet splitting follows from an enlarged Higgs sector and an appropriate vacuum alignment~\cite{InoueKakutoTakano1986,BarbieriDvaliStrumia1993,BarbieriDvaliMoretti1993,BerezhianiCsakiRandall1995,Berezhiani1995,CsakiRandall1996,ChackoMohapatra1998,ShafiTavartkiladze2001}.  Discrete $R$ symmetries can restrict the allowed superpotential operators~\cite{LeeR2011}, and a supersymmetric $331$ model from an $SU(6)$ GUT model with a gauged $Z_4^R$ symmetry was studied in Ref.~\cite{ChenLiuTeng2021}.  Here we impose the light-doublet condition as a tree-level determinant relation.  The colored-triplet inverse mass matrix is kept as an independent Higgs-sector quantity, so the dimension-five proton-decay amplitude is written in terms of the physical triplet propagator without assuming a specific triplet spectrum.

Gauge mediation relates the soft supersymmetry-breaking terms to a common spurion~\cite{DineNelson1993,DineNelsonShirman1995,DineNelsonNirShirman1996,GiudiceRattazzi1999,MeadeSeibergShih2009,MartinPrimer}.  A precision unification calculation includes both messenger and heavy-field thresholds consistently in the renormalization-group evolution~\cite{MartinVaughn1994,LangackerPolonsky1993,BaggerMatchevPierce1995,Pierce1997}.  In the selected higher-dimensional operator, the messenger spurion also generates a singlet Majorana mass, but the displayed selection rules do not exclude competing lower-dimensional operators.  The messenger mass can then be eliminated to obtain a relation between the unified coupling and the gauge-singlet lepton-number scale.  At one loop, the logarithmic coefficient depends only on the number of messenger multiplets and is independent of the unspecified flavor coefficients.

The $331$-breaking pair also defines a nonminimally coupled inflationary direction.  Nonminimal scalar inflation has been studied in both nonsupersymmetric and supergravity formulations~\cite{FutamaseMaeda1989,FakirUnruh1990,Kaiser1995,KomatsuFutamase1999,BezrukovShaposhnikov2008,BarbonEspinosa2009,BurgessLeeTrott2009,LernerMcDonald2010,KalloshLinde2010,BezrukovMagnin2011}; gauge-charged inflation associated with a TeV-scale inverse seesaw was considered in a different gauge theory in Ref.~\cite{Moursy2021}.  The scalar tilt, tensor-to-scalar ratio, and running are constrained by present CMB data and will be probed further by CMB-S4 and LiteBIRD~\cite{PlanckInflation2018,BICEP2021,ACTExtended2025,Balkenhol2025,CMBS4Science,LiteBIRD2023}.  In the restricted inflationary potential, the scalar amplitude sets the superpotential coupling that controls the radial mass in the broken $331$ vacuum.  We consequently obtain a relation between the CMB normalization, the radial spectrum, and the associated Barbieri--Giudice sensitivity.

The paper is organized as follows.  In \Sec{sec:model} we specify the unified representations, selection rules, and symmetry-breaking vacua.  In \Sec{sec:iss} we derive the inverse-seesaw null-vector relation and the leading symmetry-breaking correction.  Gauge mediation, two-loop unification, the soft spectrum, and the broken-vector mass are discussed in \Sec{sec:gauge}.  In \Sec{sec:proton} we express the proton-decay constraints in terms of the physical gauge-boson and colored-triplet parameters.  The inflationary observables and the transverse supergravity stability are evaluated in \Sec{sec:inflation}.  In \Sec{sec:feasibility} we combine the three scale relations and distinguish independent inputs from derived quantities.  The appendices contain the group-theory sums, reconstruction formulas, and intermediate derivations. 
The consistency audit in Sec.~\ref{sec:consistency} identifies symmetry-level obstructions before any phenomenological interpretation of the conditional numerical benchmarks.

\section{Supersymmetric $331$ model from the $SU(6)$ GUT model}\label{sec:model}
\subsection{Embedding and chiral matter}\label{sec:embedding}
The $SU(6)$ generator associated with the intermediate Abelian factor is normalized in the fundamental representation as
\begin{align}
T_X=\frac{1}{2\sqrt3}\diag(-1,-1,-1,+1,+1,+1),\nonumber\\
\Tr(T_X^2)=\frac12 .
\label{eq:tx}
\end{align}
Writing $q_6\equiv1/(2\sqrt3)$, the fundamental branches as
\begin{equation}
\mathbf6=(\mathbf3,\mathbf1,-q_6)\oplus(\mathbf1,\mathbf3,+q_6).
\label{eq:branch6}
\end{equation}
The antisymmetric representation is the exterior product $\mathbf{15}=\wedge^2\mathbf6$.  Its components with two color indices, two left indices, or one index of each type decompose as
\begin{equation}
\mathbf{15}=(\overline{\mathbf3},\mathbf1,-2q_6)
\oplus(\mathbf1,\overline{\mathbf3},+2q_6)
\oplus(\mathbf3,\mathbf3,0).
\label{eq:branch15}
\end{equation}
The charge embedding used below is
\begin{equation}
Q=T_{3L}+\frac{1}{\sqrt3}T_{8L}+\frac{2}{\sqrt3}X .
\label{eq:charge}
\end{equation}
With the normalization in \Eq{eq:tx}, this embedding gives the GUT-normalized matching relation used in \Sec{sec:gauge}.  The charge assignment is the $SU(6)$ realization of the original and right-handed-neutrino $331$ constructions~\cite{PisanoPleitez1992,Frampton1992,Montero1993,FootLongTran1994,Long1996,CorianoMelle2024}.  The representation branching follows the $331\subset SU(6)$ embedding of Refs.~\cite{Deppisch2016,Le2020}.

Each chiral family is assigned to
\begin{equation}
\mathcal F_i=\mathbf{15}_i\oplus\overline{\mathbf6}_i
\oplus\overline{\mathbf6}'_i,
\qquad i=1,2,3,
\label{eq:family}
\end{equation}
We also introduce a gauge singlet $N_{s i}$.  With the conventional cubic anomaly coefficients $A(\mathbf6)=1$ and $A(\mathbf{15})=2$, the anomaly of one family is
\begin{equation}
A(\mathcal F_i)=A(\mathbf{15})+2A(\overline{\mathbf6})=2-1-1=0.
\label{eq:anomaly}
\end{equation}
Thus the continuous $SU(6)^3$ gauge anomaly cancels within each family.  The Higgs representations are vectorlike under $SU(6)$, and their higgsino contributions cancel between conjugate pairs.

\begin{table*}[t]
\centering
\caption{Superfields relevant to the scale relations.  The $G_{331}$ column lists the components used below; conjugate components are understood for the vectorlike Higgs and messenger pairs.  Here $R$ denotes the $Z_4^R$ charge and $P_\nu$ the $Z_2^\nu$ parity.  The ordinary and primed antifundamentals cannot mix while this parity is unbroken.}
\label{tab:fields}
\scriptsize
\resizebox{\textwidth}{!}{%
\begin{tabular}{@{}llllll@{}}
\toprule
superfield & $SU(6)$ & representative $G_{331}$ content & $R$ & $P_\nu$ & role \\
\midrule
$15_i$ & $\mathbf{15}$ & $(\mathbf3,\mathbf3,0)\oplus(\overline{\mathbf3},\mathbf1,-1/\sqrt3)\oplus(\mathbf1,\overline{\mathbf3},+1/\sqrt3)$ & $1$ & $+$ & quarks, charged leptons, $\nu^c$ components \\
$\bar6_i$ & $\overline{\mathbf6}$ & $(\overline{\mathbf3},\mathbf1,+q_6)\oplus(\mathbf1,\overline{\mathbf3},-q_6)$ & $1$ & $+$ & ordinary antifundamental matter \\
$\bar6'_i$ & $\overline{\mathbf6}$ & $(\overline{\mathbf3},\mathbf1,+q_6)\oplus(\mathbf1,\overline{\mathbf3},-q_6)$ & $1$ & $-$ & inverse-seesaw lepton multiplet \\
$N_{s i}$ & $\mathbf1$ & $(\mathbf1,\mathbf1,0)$ & $3$ & $-$ & inverse-seesaw singlet \\
$H_{15}+\bar H_{15}$ & $\mathbf{15}+\overline{\mathbf{15}}$ & electroweak triplet/antitriplet fragments & $0,2$ & $+$ & up-type and neutrino operators \\
$H_6+\bar H_6$ & $\mathbf6+\overline{\mathbf6}$ & electroweak triplet/antitriplet fragments & $2,0$ & $+$ & down-type operators \\
$\Phi+\bar\Phi$ & $\mathbf6+\overline{\mathbf6}$ & $(\mathbf1,\mathbf3,+q_6)\oplus(\mathbf1,\overline{\mathbf3},-q_6)$ & $2,2$ & $+$ & $331$ breaking and inflation \\
$S$ & $\mathbf1$ & $(\mathbf1,\mathbf1,0)$ & $2$ & $+$ & $F$-term breaking/inflation singlet \\
$A,B$ & $\mathbf{35}$ & adjoint fragments & $0,2$ & $+$ & $SU(6)\to G_{331}$ vacuum \\
$Z$ & $\mathbf1$ & $(\mathbf1,\mathbf1,0)$ & $2$ & $+$ & adjoint constraint \\
$\Sigma_L,\Sigma_C$ & $\mathbf{35}$ fragments & $(\mathbf1,\mathbf8,0)$, $(\mathbf8,\mathbf1,0)$ & --- & $+$ & intermediate gauge thresholds \\
$X$ & $\mathbf1$ & spurion & $2$ & $+$ & $\langle X\rangle=M_{\rm GM}+\theta^2F_X$ \\
$\Psi_A+\bar\Psi_A$ & $\mathbf6+\overline{\mathbf6}$ & complete messenger multiplet & $0,0$ & $+$ & gauge mediation, $A=1,2$ \\
\bottomrule
\end{tabular}%
}
\end{table*}

\subsection{Selection rules and Higgs mixing}\label{sec:selection}
The superpotential carries $Z_4^R$ charge two.  We assign
\begin{align}
R(15_i,\bar6_i,\bar6'_i)&=1, & R(N_s)&=3,\nonumber\\
R(H_{15},\bar H_6)&=0, & R(\bar H_{15},H_6)&=2,\nonumber\\
R(\Phi,\bar\Phi,S)&=2.&&
\label{eq:rcharges}
\end{align}
The Higgs multiplets and their vectorlike conjugates carry different discrete charges, as allowed for conjugate gauge representations.  The assignments in \Eq{eq:rcharges} allow the Yukawa and Higgs-mixing operators in \Eqs{eq:superpotential}{eq:whiggs} and distinguish the two fields in each vectorlike pair.  We use $Z_4^R\times Z_2^\nu\times Z_3^G$ as selection rules of the effective unified theory, without specifying their higher-scale origin.  Discrete $R$ symmetries and the associated operator restrictions have been studied in supersymmetric models~\cite{LeeR2011,ChenLiuTeng2021}.  Under $Z_2^\nu$, the fields $\bar6'_i$ and $N_{s i}$ are odd.  The required renormalizable terms in the matter and $331$ sectors are
\begin{align}
W={}&\frac14y^u_{ij}15_i15_jH_{15}
+\sqrt2y^d_{ij}15_i\bar6_j\bar H_6\nonumber\\
&+\frac14y^\nu_{ij}\bar6'_i\bar6'_jH_{15}
+Y_{ij}\bar6'_i\Phi N_{s j}\nonumber\\
&+\kappa S\!\left(\bar\Phi\Phi-\frac{v_0^2}{2}\right)
+W_H+W_G+W_{\rm GM}.
\label{eq:superpotential}
\end{align}
Here $v_0$ is the mass parameter in the superpotential.  It differs from the physical $331$ vacuum expectation value because the gauge-mediated soft terms shift the minimum.  The relation between these two scales is derived in \Sec{sec:gauge}.

The Higgs mixing terms are
\begin{align}
W_H={}&\mu_{15}\bar H_{15}H_{15}+\mu_6\bar H_6H_6\nonumber\\
&+\lambda_HH_{15}\bar H_6\bar\Phi
+\bar\lambda_H\bar H_{15}H_6\Phi .
\label{eq:whiggs}
\end{align}
After $\langle\Phi\rangle=\langle\bar\Phi\rangle=v_{331}/\sqrt2$, the two electroweak doublet pairs have
\begin{equation}
\mathcal M_D=
\begin{pmatrix}
\mu_{15}&\lambda_Hv_{331}/\sqrt2\\
\bar\lambda_Hv_{331}/\sqrt2&\mu_6
\end{pmatrix}.
\label{eq:doubletmatrix}
\end{equation}
A light supersymmetric doublet pair requires
\begin{equation}
\det\mathcal M_D=
\mu_{15}\mu_6-\frac12\lambda_H\bar\lambda_Hv_{331}^2=0,
\qquad
\bar\lambda_H=\frac{2\mu_{15}\mu_6}{\lambda_Hv_{331}^2}.
\label{eq:doubletcondition}
\end{equation}
For $(\mu_{15},\mu_6,\lambda_H)=(7.00,6.00,0.200)$ in TeV units and $v_{331}=30.0$ TeV, \Eq{eq:doubletcondition} gives $\bar\lambda_H=0.467$.  A direct singular-value decomposition gives one vanishing supersymmetric eigenvalue and a heavy doublet mass $M_{D_H}=14.2$ TeV.  This heavy eigenvalue is included in the gauge matching below.  Equation~\eqref{eq:doubletcondition} represents a tree-level tuning in the two-pair Higgs sector.  A dynamical treatment of doublet--triplet splitting requires an extended Higgs sector.  Pseudo-Goldstone and missing-VEV mechanisms provide such realizations in supersymmetric $SU(6)$~\cite{InoueKakutoTakano1986,BarbieriDvaliStrumia1993,BarbieriDvaliMoretti1993,BerezhianiCsakiRandall1995,Berezhiani1995,CsakiRandall1996,ChackoMohapatra1998,ShafiTavartkiladze2001}.  We retain the minimal two-pair sector needed for the present calculation; consequently, the colored-triplet propagator entering dimension-five proton decay remains an independent quantity.

The gauge contraction in the neutrino operator imposes an additional algebraic restriction.  Since the chiral superfields commute and $H_{15}^{ab}=-H_{15}^{ba}$,
\begin{equation}
y^{\nu T}=-y^\nu .
\label{eq:antisymmetry}
\end{equation}
This antisymmetry determines the leading neutrino rank discussed in \Sec{sec:iss}.

Gauge mediation is generated by two complete messenger pairs through
\begin{align}
W_{\rm GM}&=\sum_{A=1}^{2}\lambda_A X\Psi_A\bar\Psi_A,\nonumber\\
\langle X\rangle&=M_{\rm GM}+\theta^2F_X,
\qquad
\Lambda_X\equiv\frac{F_X}{M_{\rm GM}}.
\label{eq:wgm}
\end{align}
Above the messenger threshold, complete $\mathbf6+\overline{\mathbf6}$ multiplets leave the relative one-loop beta-function coefficients unchanged but modify the common unified coupling.  Their contribution is evaluated in \Sec{sec:gauge}.

\subsection{Unified and 331 vacua}\label{sec:vacua}
The adjoint sector is defined by
\begin{align}
W_G={}&\lambda_ZZ\left(\Tr A^2-\frac{v_A^2}{2}\right)\nonumber\\
&+\lambda_B\Tr\!\left[B\left(A^2-\frac{\Tr A^2}{6}\mathbf1_6\right)\right].
\label{eq:wg}
\end{align}
We assign a non-$R$ $Z_3^G$ charge of one to $A$, $B$, and $Z$, a charge of two to the $v_A^2$ spurion, and zero charge to the light matter and Higgs fields.  On the branch $B=Z=0$, the $F_B$ and $F_Z$ conditions give
\begin{equation}
\langle A\rangle=\frac{v_A}{2\sqrt3}
\diag(-1,-1,-1,+1,+1,+1),
\label{eq:avev}
\end{equation}
This vacuum leaves $G_{331}=SU(3)_C\times SU(3)_L\times U(1)_X$ unbroken.  The $Z_3^G$ assignment is also used in the neutrino-operator analysis of \Sec{sec:iss}.

The superpotential in \Eq{eq:wg} determines the displayed vacuum but does not by itself provide a mass for every physical component of $B$.  In particular, the $B$ bifundamentals are not Goldstone multiplets, whereas the corresponding components of $A$ are eaten by the broken gauge vectors.  A complete ultraviolet Higgs sector requires additional mass operators or adjoint multiplets that lift these states and the unwanted colored partners of $H_6$, $H_{15}$, and $\Phi$.  We do not infer those mass matrices from \Eq{eq:wg}.  The GUT-scale threshold calculation below uses an explicitly declared conditional heavy spectrum; its crossing is not presented as a prediction of the minimal superpotential.

The lower breaking direction is the real $D$-flat trajectory
\begin{equation}
\Phi=\bar\Phi=\frac{\varphi}{2},\qquad S=0,
\label{eq:dflat}
\end{equation}
The physical minimum is $\langle\varphi\rangle=\sqrt2v_{331}$, and we use $v_{331}=3.00\times10^4\,{\rm GeV}$ throughout.  The split adjoint components have holomorphic masses
\begin{equation}
W_R=m_R\bar RR+\lambda_R\bar RAR,
\qquad M_{R,r}=m_R+\lambda_Rc_rv_A,
\label{eq:splitmass}
\end{equation}
Here $c_r$ is the eigenvalue associated with the component $r$.  We take the color-octet mass to be $M_{\Sigma_C}=1.00\times10^{12}\,{\rm GeV}$ and determine the weak-octet mass from gauge-coupling unification.

The illustrative split-multiplet formula in Eq.~\eqref{eq:splitmass} also requires a charged mass spurion under the stated $Z_3^G$. For $q_G(A)=1$, simultaneous invariance of $m_R\bar RR$ and $\lambda_R\bar RAR$ requires $q_G(\bar R)+q_G(R)=2$ and $q_G(m_R)=1$ modulo three. A neutral numerical parameter $m_R$ is inconsistent with the original assignment. In the conditional threshold model we therefore interpret $m_R$ as the expectation value of a charge-one spurion. The ultraviolet origin and its spectrum are not specified.

\section{Selection-rule consistency and necessary completion}\label{sec:consistency}
\subsection{Ordinary-lepton mixing and colored partners}
The parity $P_\nu$ in Table~\ref{tab:fields} is even for $\bar6_i$ and odd for $\bar6'_i$ and $N_{si}$. With the parity-even vacuum, the ordinary active-neutrino field $\nu_{L,e}\subset\bar6$ and the neutral fields in $\bar6'$ belong to distinct parity sectors. In particular, the gauge-invariant coupling $\bar6_i\Phi N_{sj}$ is odd. A schematic neutral-fermion mass matrix in the parity-ordered basis has the structure
\begin{equation}
{\cal M}_{\rm neutral}=\begin{pmatrix}{\cal M}_{+}&0\\0&{\cal M}_{-}\end{pmatrix},\qquad
\left(\nu^{}_{L,e}\right)\not\leftrightarrow (\nu'_{L},\nu^{\prime c},N_s).
\label{eq:parityblock}
\end{equation}
Consequently, the inverse-seesaw matrix of Eq.~\eqref{eq:fullmass} cannot be identified with the mass matrix of the observed active neutrinos on the stated unbroken-parity vacuum. This is a strict selection-rule obstruction, not a small perturbative correction. The quoted $m_D$ fit below is an algebraic example within the odd sector or in a separately completed active-coupled theory.

The Yukawa $15_i\bar6'_j\bar H_6$ has $R$ charge two but odd $P_\nu$ and is excluded. The allowed $15_i\bar6_j\bar H_6$ does not pair the corresponding primed exotic colored fermions. A phenomenologically viable theory must provide an explicit massive colored-fermion pairing matrix before their contributions are omitted from the beta functions. A lower-energy numerical unification crossing is not a substitute for this mass calculation.

For illustration, a parity-odd gauge-singlet spurion $P_0$ with $R(P_0)=0$ permits
\begin{equation}
\Delta W_\nu=\frac{h_{ij}}{M_0}P_0\bar6_i\bar6'_jH_{15},\qquad
m_{\rm mix}=\frac{v_u\langle P_0\rangle}{\sqrt2M_0}h.
\label{eq:parityspurion}
\end{equation}
An independent parity-odd spurion $P_2$ with $R(P_2)=2$ permits a potentially heavy exotic pairing operator
\begin{equation}
\Delta W_{\rm col}=\frac{g_{ij}}{M_2}P_2\,15_i\bar6'_j\bar\Phi,
\qquad m_{\rm col}\sim g\frac{\langle P_2\rangle v_{331}}{\sqrt2M_2}.
\label{eq:colorspurion}
\end{equation}
The contractions in Eqs.~\eqref{eq:parityspurion} and \eqref{eq:colorspurion} are gauge invariant and carry the required $Z_4^R$ charge. They are examples of necessary new interactions, not operators of the original parity-unbroken model. Their introduction changes the physical neutral and exotic mass matrices and can spoil a purely antisymmetric active Dirac texture. A realistic flavor construction must therefore derive the light eigenstates, charged-fermion spectrum and the texture constraints simultaneously. No complete such fit is asserted here.

\subsection{Allowed singlet masses and omitted couplings}
The proposed inverse-seesaw singlet has $R(N_s)=3$. The superpotential selection rule $R(W)=2\pmod4$ allows $N_{si}N_{sj}$ because $3+3=6=2\pmod4$. Both singlets are odd under $P_\nu$, so the product is even. With the stated neutral $Z_3^G$ assignment, the term is allowed by every displayed symmetry:
\begin{equation}
\Delta W_{\rm allowed}\supset \frac12(M_S)_{ij}N_{si}N_{sj}.
\label{eq:allowedmajorana}
\end{equation}
Similarly, $K\supset d_{ij}X^\dagger N_{si}N_{sj}/(2M_K)+\mathrm{h.c.}$ carries zero $Z_4^R$ charge and is allowed. Its contribution to the holomorphic singlet mass is of order
\begin{equation}
\delta\mu_S\sim \frac{dF_X}{M_K},\qquad
\frac{\delta\mu_S}{\mu_0}\sim\frac{2d}{c}\frac{M_{\rm LNV}^3}{M_Kv_{331}^2}.
\label{eq:mulower}
\end{equation}
The higher-dimensional operator in Eq.~\eqref{eq:kahleriss} is therefore not the leading symmetry-allowed source. At the input $F_X=9.00\times10^{20}\,{\rm GeV}^2$, taking $M_K=M_{\rm LNV}=4.09\times10^{11}$ GeV yields $F_X/M_K=2.20\times10^9$ GeV. Even Planck suppression gives $F_X/M_P=370$ GeV, compared with the selected $\mu_0=5.92$ keV. Without an additional ultraviolet selection rule or an explicitly demonstrated cancellation, the quoted keV mass and the derived messenger--lepton-number scale relation are conditional parameterizations rather than natural predictions.

The following gauge- and $Z_4^R$-allowed terms are also even under $P_\nu$ and neutral under the stated $Z_3^G$:
\begin{equation}
\Delta W\supset m_{6\Phi}\bar H_6\Phi +\lambda_{X\Phi}X\bar\Phi\Phi
+\frac{\lambda_{S3}}3S^3+\lambda_{SH}SH_6\bar\Phi.
\label{eq:omittedterms}
\end{equation}
Here $m_{6\Phi}$ has mass dimension one and all $\lambda$ coefficients are dimensionless. These operators cannot be dropped on the basis of the displayed discrete symmetries. In particular, the $\lambda_{X\Phi}$ term gives $F_{\Phi}\supset\lambda_{X\Phi}X\bar\Phi$ and alters both the proposed broken vacuum and the inflationary trajectory. At the benchmark, retaining a supersymmetric mixing scale smaller than $m_\rho=127$ GeV requires $|\lambda_{X\Phi}|\lesssim4.23\times10^{-14}$ for $|\langle X\rangle|=3.00\times10^{15}$ GeV. The stronger condition is a phenomenological tolerance, not a consequence of the presented symmetries.

\begin{figure*}[t]
\centering
\includegraphics[width=0.485\textwidth]{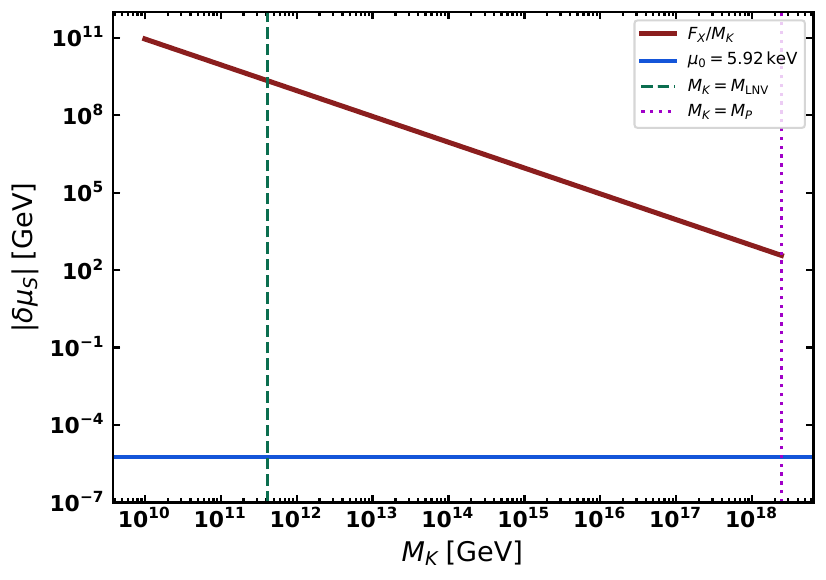}\hfill
\includegraphics[width=0.485\textwidth]{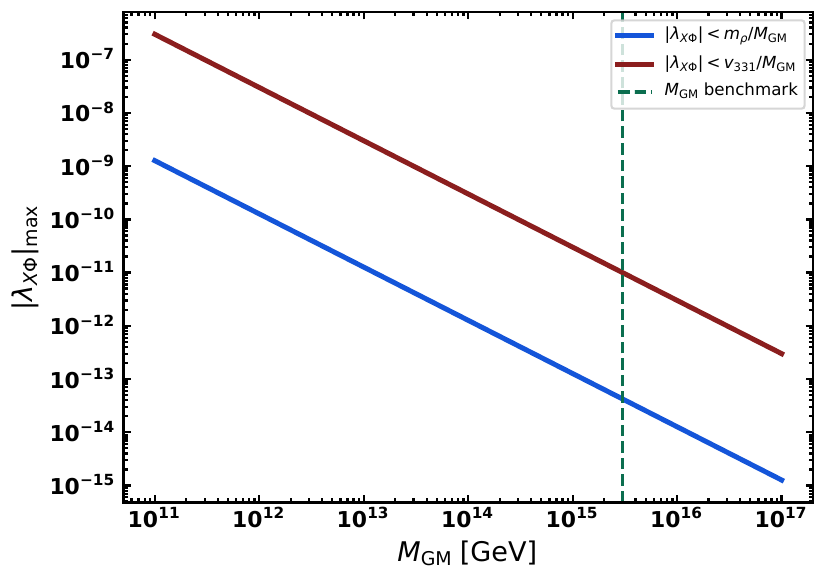}
\caption{Quantitative consistency diagnostics at the stated benchmark. Left: size of the allowed lower-dimensional $X^\dagger N_sN_s$ K\"ahler contribution for unit Wilson coefficient, compared with the selected $5.92$ keV mass, as a function of its suppression scale $M_K$. Right: upper limits on the independent $X\bar\Phi\Phi$ coupling imposed by requiring $|\lambda_{X\Phi}M_{\rm GM}|$ to remain below either the radial mass or the $331$ scale. These are operator-level naturalness tests, not fits of a completed ultraviolet theory.}
\label{fig:operatoraudit}
\end{figure*}

Figure~\ref{fig:operatoraudit}(a) compares the inverse-mass scaling of the symmetry-allowed $X^\dagger N_sN_s/M_K$ contribution with the chosen $\mu_0=5.92$ keV.  For unit Wilson coefficient its value remains $370$ GeV even at $M_K=M_P$.  Figure~\ref{fig:operatoraudit}(b) shows the distinct bound $|\lambda_{X\Phi}|\lesssim m_\rho/M_{\rm GM}$ and the weaker bound involving $v_{331}$.  Neither curve represents a symmetry-enforced suppression.  Thus the benchmark scale relation requires additional operator selection beyond the displayed charge assignments.

\subsection{Adjoint threshold and scope of the inflationary calculation}
For $\langle A\rangle=a\operatorname{diag}(-\mathbf1_3,+\mathbf1_3)$, the quadratic $B$--$A$ mixing from $\operatorname{Tr}(BA^2)$ is proportional to $\{\langle A\rangle,\delta A\}$. In the off-diagonal color--left blocks, the anticommutator vanishes exactly because $-a+a=0$. The $18$ complex bifundamental components of $B$ therefore receive no bilinear superpotential mass from Eq.~\eqref{eq:wg}; unlike the corresponding $A$ directions they are not eaten. Their masses and the heavy-field contribution to the threshold running require additional interactions. The benchmark $(M_G,\alpha_G)$ consequently belongs to a \emph{postulated} threshold spectrum, not the mass spectrum calculated from the displayed $W_G$.

\begin{figure}[t]
\centering
\includegraphics[width=0.97\columnwidth]{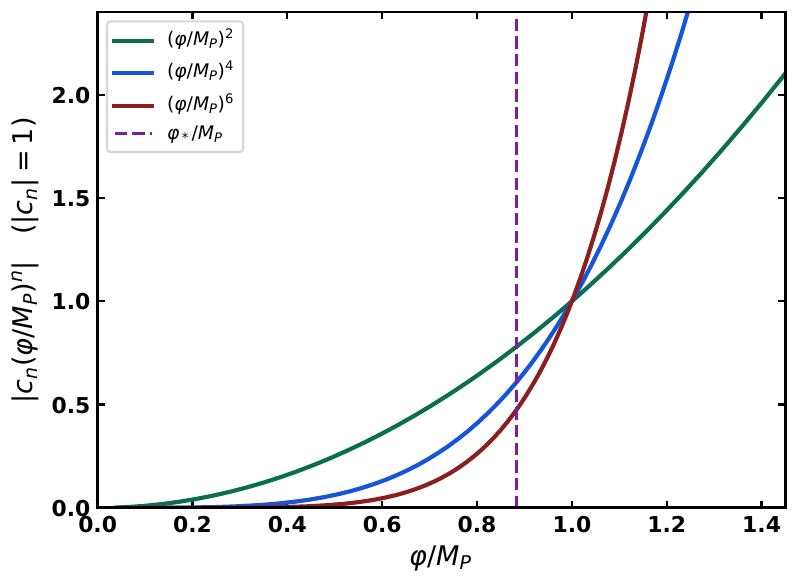}
\caption{Magnitude of the simple large-field expansion factors $(\varphi/M_P)^n$ for $n=2,4,6$, with unit Wilson coefficient; the line marks $\varphi_*/M_P=0.883$. These factors illustrate the absence of a generic small-field expansion, not a computed correction to the specific inflationary potential.}
\label{fig:eftpower}
\end{figure}

The large-field trajectory has $\varphi_*/M_P=0.883$ and $\varphi_*/M_G\simeq48$ for the benchmark. Generic unprotected operators $c_n(\varphi/M_P)^n$ are not suppressed parametrically for small $n$ at horizon exit. The neutral $\Phi,\bar\Phi,S$ covariant Hessian does not determine stability of all charged, colored, adjoint, Higgs and messenger scalars along this trajectory, nor does $H_*/M_G\ll1$ establish that these fields remain fixed while $\varphi$ changes. A full multifield and ultraviolet operator analysis is necessary before the quoted CMB observables are promoted to complete-model predictions.
The power counting in Fig.~\ref{fig:eftpower} makes the large-field limitation explicit. At $\varphi_*/M_P=0.883$, the factors $(\varphi_*/M_P)^2$, $(\varphi_*/M_P)^4$, and $(\varphi_*/M_P)^6$ are approximately $0.780$, $0.608$, and $0.474$. These numbers are coefficients of possible operators, not shifts calculated from a complete K\"ahler potential.

\section{Rank-two inverse seesaw and symmetry-breaking corrections}\label{sec:iss}
For a neutral subsector assumed to couple to the observed active leptons after the additional parity-breaking interactions of Sec.~\ref{sec:consistency}, the usual inverse-seesaw matrix in the basis $(\nu_L,\nu^c,N_s)$ is
\begin{align}
\mathcal M_\nu&=
\begin{pmatrix}
0&m_D&0\\m_D^T&0&M_N\\0&M_N^T&\mu_S
\end{pmatrix},
\label{eq:fullmass}\\
m_D&=\frac{v_u}{\sqrt2}y^\nu,
\qquad
M_N=\frac{v_{331}}{\sqrt2}Y.
\label{eq:mdmn}
\end{align}
For $\|\mu_S\|\ll\|M_N\|$, block diagonalization gives the inverse-seesaw relation~\cite{Mohapatra1986,DeppischValle2005,Dias2012,DevMohapatra2010,ParkWang2011,Boucenna2014}
\begin{equation}
m_\nu=m_DM_N^{-T}\mu_SM_N^{-1}m_D^T.
\label{eq:lightmass}
\end{equation}
A nonvanishing antisymmetric $3\times3$ matrix has rank two.  It follows from \Eq{eq:antisymmetry} that
\begin{equation}
\operatorname{rank}(m_D)=2,
\qquad
\operatorname{rank}(m_\nu)\le2,
\qquad
\det m_\nu=0
\label{eq:rank}
\end{equation}
Thus, at leading order, one light-neutrino mass vanishes.  The same rank condition occurs in earlier $331$ inverse-seesaw constructions~\cite{Dias2012} and in the nonsupersymmetric $331$ model from the $SU(6)$ GUT model of Ref.~\cite{Le2020}.  We next determine how symmetry-allowed operators in the supersymmetric theory lift the zero eigenvalue.

The lepton-number-violating matrix is generated after $331$ breaking by the effective K\"ahler operator
\begin{equation}
K_{\rm LNV}\supset
\frac{c_{ij}}{2M_{\rm LNV}^3}
X^\dagger(\bar\Phi\Phi)N_{s i}N_{s j}+{\rm h.c.}
\label{eq:kahleriss}
\end{equation}
The scale $M_{\rm LNV}$ denotes the mass scale of the gauge-singlet mediators and is distinct from the charged gauge-messenger mass $M_{\rm GM}$.  With \Eq{eq:wgm}, we obtain
\begin{equation}
\mu_S=\mu_0c,
\qquad
\mu_0=\frac{F_Xv_{331}^2}{2M_{\rm LNV}^3}
=\frac{\Lambda_XM_{\rm GM}v_{331}^2}{2M_{\rm LNV}^3}.
\label{eq:mus}
\end{equation}
Hence
\begin{equation}
M_{\rm LNV}=
\left(\frac{\Lambda_XM_{\rm GM}v_{331}^2}{2\mu_0}\right)^{1/3}.
\label{eq:mlnv}
\end{equation}
This algebraic relation applies only if the lower-dimensional operators in Eq.~\eqref{eq:mulower} and the direct Majorana term in Eq.~\eqref{eq:allowedmajorana} are suppressed by an additional mechanism not contained in the displayed charge table.  For an illustrative texture reconstruction, not a fit of the full parity-unbroken theory, we use normal ordering and set $m_D=10.0A\,{\rm GeV}$.  The antisymmetric matrix $A$ has unit Frobenius norm, and its null vector is identified with the massless PMNS eigenvector.  The rank-two condition by itself permits either mass ordering; normal ordering is used only for the numerical reconstruction.  We take the heavy masses to be $(2.00,2.50,3.00)$ TeV.  A Moore--Penrose reconstruction restricted to the rank-two image gives
\begin{equation}
\mu_0=5.92~{\rm keV},
\qquad
(m_1,m_2,m_3)=(0,8.61,50.2)~{\rm meV},
\label{eq:issnumbers}
\end{equation}
The relative reconstruction residual is $2.74\times10^{-16}$.  For $\Lambda_X=3.00\times10^5\,{\rm GeV}$ and the gauge-messenger scale specified below, \Eq{eq:mlnv} gives $M_{\rm LNV}=4.09\times10^{11}\,{\rm GeV}$.  The heavy-light mixing matrix $\Theta=m_DM_N^{-1}$ gives the heavy-neutral-lepton contribution
\begin{equation}
\BR(\mu\to e\gamma)_{\rm HNL}
=\frac{3\alpha}{32\pi}
\left|(\Theta\Theta^\dagger)_{e\mu}\right|^2
=3.40\times10^{-16},
\label{eq:mueg}
\end{equation}
This value is below the MEG II limit~\cite{MEGII2025}.  In low-scale seesaw models, the heavy-neutral-lepton and supersymmetric flavor amplitudes require separate treatment~\cite{IlakovacPilaftsis2009,IlakovacPilaftsisPopov2013,Abada2014,Vicente2015,KriewaldTeixeira2025}.  Within this illustrative inverse-seesaw block, the reconstruction gives $\max|\Theta_{\alpha i}|=2.33\times10^{-3}$, $\max|\eta_{\alpha\alpha}|=3.43\times10^{-6}$, and $\max_{\alpha\ne\beta}|\eta_{\alpha\beta}|=1.64\times10^{-6}$, where $\eta\equiv\Theta\Theta^\dagger/2$.  The conditional branching fraction is a consequence of the multi-TeV values of $M_N$ and the $10$ GeV Dirac scale, for which $\Theta\sim m_DM_N^{-1}$.  Supersymmetric loop contributions depend on additional slepton flavor parameters and are not included in \Eq{eq:mueg}.

The renormalization-group evolution of the neutrino parameters depends on the charged-lepton and neutrino Yukawa matrices and on the successive heavy thresholds~\cite{Antusch2005,ParkWang2011}.  The reconstructed neutrino spectrum is hierarchical rather than quasi-degenerate and consequently does not receive the mass-degeneracy enhancement of the running.  Numerical shifts of the mixing angles depend on $\tan\beta$ and on the complete charged-lepton and slepton flavor sectors, which are kept as independent inputs.  By contrast, the rank condition and determinant scaling derived below are basis-independent algebraic results.  A precision calculation of the low-energy mixing angles requires a complete flavor fit.

We now consider higher-dimensional chiral operators that correct the rank-two limit.  Adjoint insertions preserve the leading structure: in the vacuum of \Eq{eq:avev}, every polynomial in $A$ is proportional to the identity within the unbroken $SU(3)_L$ block and retains the antisymmetric contraction of the light neutral fields.  Among the superpotential operators constructed from the displayed chiral multiplets, the lowest-dimensional $331$-breaking term that can generate a symmetric flavor component is
\begin{equation}
\Delta W_D=
\frac{d_{ij}}{M_{\rm UV}^2}
\left(\bar6'_{ia}H_{15}^{ac}\bar\Phi_c\right)
\left(\bar6'_{jb}\Phi^b\right),
\label{eq:phirankoperator}
\end{equation}
Here $M_{\rm UV}$ is the scale suppressing the operator.  Since a chiral superfield has mass dimension one and the superpotential has dimension three, the product of five chiral superfields in \Eq{eq:phirankoperator} requires the coefficient $M_{\rm UV}^{-2}$.  The two $\bar6'$ fields enter through different gauge contractions in \Eq{eq:phirankoperator}.  After $\Phi$ and $\bar\Phi$ develop their $331$-breaking expectation values, the resulting flavor matrix may consequently contain a symmetric component.  This operator is analogous to the higher-dimensional $SU(6)$ corrections used to perturb antisymmetric neutrino textures in Ref.~\cite{ChackoDevMohapatraThapa2020}, but here it contributes to the inverse-seesaw mass formula.

Writing
\begin{equation}
m_D=m_D^{(0)}+\epsilon_\Phi\,\delta m_D,
\qquad
\epsilon_\Phi\sim\frac{v_{331}^2}{M_{\rm UV}^2},
\label{eq:rankpert}
\end{equation}
For the rank-two matrix $m_D^{(0)}$, Jacobi's determinant expansion gives
\begin{equation}
\det(m_D)=\epsilon_\Phi\,
\Tr\!\left[\operatorname{adj}(m_D^{(0)})\,\delta m_D\right]
+\mathcal O(\epsilon_\Phi^2).
\label{eq:detmdexp}
\end{equation}
The adjugate of a rank-two $3\times3$ matrix has rank one and projects onto the null directions.  If the correction has a nonzero projection on these directions, then $\det m_D=\mathcal O(\epsilon_\Phi)$.  The determinant of \Eq{eq:lightmass} is
\begin{equation}
\det m_\nu=
\frac{[\det(m_D)]^2\det(\mu_S)}{[\det(M_N)]^2}.
\label{eq:detmnu}
\end{equation}
For finite $m_2$ and $m_3$, the Takagi relation $|\det m_\nu|=m_{\rm lightest}m_2m_3$ then gives
\begin{equation}
m_{\rm lightest}=\mathcal O(\epsilon_\Phi^2)m_{\rm atm}
=\mathcal O\!\left[\left(\frac{v_{331}}{M_{\rm UV}}\right)^4\right]m_{\rm atm}.
\label{eq:fourthsuppression}
\end{equation}

Let $r$ and $\ell$ be right and left null vectors of $m_D^{(0)}$,
\begin{equation}
m_D^{(0)}r=0,\qquad \ell^Tm_D^{(0)}=0,
\qquad
\operatorname{adj}(m_D^{(0)})={\cal C}\,r\ell^T,
\label{eq:nulladj}
\end{equation}
Here ${\cal C}$ has mass dimension two.  Substituting \Eq{eq:nulladj} into Jacobi's formula gives the leading determinant and its coefficient:
\begin{equation}
\det m_D=\epsilon_\Phi {\cal C}\,
\ell^T\delta m_D r+\mathcal O(\epsilon_\Phi^2).
\label{eq:detmdnull}
\end{equation}
For nonsingular $M_N$ and $\mu_S$, and finite nonzero $m_2,m_3$, the Takagi masses obey
\begin{equation}
m_{\rm lightest}=
\epsilon_\Phi^2
\frac{|{\cal C}|^2|\ell^T\delta m_Dr|^2|\det\mu_S|}
{|\det M_N|^2m_2m_3}
+\mathcal O(\epsilon_\Phi^3).
\label{eq:mlightcoefficient}
\end{equation}
Equation~\eqref{eq:mlightcoefficient} gives the model-dependent coefficient that is not fixed by power counting.  For $\ell^T\delta m_Dr\neq0$, the lightest mass is quartically suppressed in $v_{331}/M_{\rm UV}$.  If a symmetry or flavor alignment sets this projection to zero, the first nonvanishing contribution appears at higher order.

\section{Gauge mediation and threshold unification}\label{sec:gauge}
We include the charged-messenger threshold explicitly in the gauge-coupling evolution.  Precision supersymmetric unification depends on the weak-scale, messenger, and unified thresholds~\cite{Amaldi1991,LangackerLuo1991,LangackerPolonsky1993,RossRoberts1992,BaggerMatchevPierce1995,Pierce1997}.  For an $N=1$ product gauge theory, and neglecting superpotential contributions to the gauge beta functions, the gauge terms through two loops are~\cite{MartinVaughn1994}
\begin{align}
b_a={}&\sum_iT_a(i)-3C_a(G),
\label{eq:bone}\\
B_{ab}={}&4\sum_iT_a(i)C_b(i)\nonumber\\
&+\delta_{ab}\!\left[2C_a(G)\sum_iT_a(i)-6C_a(G)^2\right].
\label{eq:btwo}
\end{align}
Here the Dynkin index $T_a(i)$ includes multiplicities under the spectator gauge groups, and $C_a(i)$ is the quadratic Casimir.  Summing the contributions of the superfields in \Tab{tab:fields} below the split-adjoint thresholds gives
\begin{equation}
b=(12,3,0),
\qquad
B=\begin{pmatrix}
12&40&24\\
5&82&24\\
3&24&48
\end{pmatrix},
\label{eq:coefficients}
\end{equation}
The ordering of the gauge factors is $(X,3L,3C)$.  The weak and color octets contribute
\begin{equation}
\Delta b_{\Sigma_L}=(0,3,0),
\qquad
\Delta b_{\Sigma_C}=(0,0,3),
\label{eq:octetb}
\end{equation}
A complete $\mathbf6+\overline{\mathbf6}$ messenger pair gives
\begin{equation}
\Delta b_{\rm mess}=(1,1,1).
\label{eq:messengerb}
\end{equation}
Since the three entries in \Eq{eq:messengerb} are equal, a complete messenger multiplet preserves the relative one-loop crossing and increases the unified gauge coupling.

At the physical $331$ scale, \Eq{eq:charge} gives
\begin{equation}
\alpha_1^{-1}=\frac15\left(4\alpha_X^{-1}+\alpha_{3L}^{-1}\right),
\qquad
\alpha_2=\alpha_{3L},
\qquad
\alpha_3=\alpha_{3C}.
\label{eq:matching}
\end{equation}
The relation $\alpha_2=\alpha_{3L}$ follows because $SU(2)_L$ is the upper-left $SU(2)$ subgroup of $SU(3)_L$.  Both groups use the canonical normalization $\Tr(T^aT^b)=\delta^{ab}/2$, so no additional embedding index is required.
The two-loop gauge evolution is
\begin{equation}
\frac{dg_a}{d\ln\mu}=
\frac{b_ag_a^3}{16\pi^2}
+\frac{g_a^3}{(16\pi^2)^2}\sum_bB_{ab}g_b^2.
\label{eq:rge}
\end{equation}
A chiral state of mass $m_r$ that is absent from the effective-theory running below a matching scale $M$ contributes
\begin{equation}
\delta\alpha_i^{-1}(M)=\frac{b_i^{(r)}}{2\pi}\ln\frac{m_r}{M}.
\label{eq:finitematching}
\end{equation}
The heavy electroweak-doublet pair contributes $\Delta b_{\rm SM}=(3/5,1,0)$.  Using \Eq{eq:matching}, this becomes $\Delta b_{331}=(1/2,1,0)$; hence its correction at $v_{331}$ is negative because $M_{D_H}<v_{331}$.  At the GUT scale we use the conditional vectors
\begin{equation}
\Delta b_H=(1/2,0,1),\quad
\Delta b_{\rm bif}=(6,3,3),\quad
\Delta b_{\rm adj}=(0,3,3),
\label{eq:heavyvectors}
\end{equation}
with mass ratios $(0.70,1.35,0.85)$ relative to $M_G$.  The first vector is that of one complete color-triplet pair in a $\mathbf6+\overline{\mathbf6}$; the previously used Abelian entry $1/6$ did not correspond to such a vectorlike pair.  The remaining vectors parametrize a bifundamental pair and diagonal adjoint fragments.  These three masses are inputs representing a heavy-sector completion and are not eigenvalues derived from \Eq{eq:wg}.  The numerical evolution includes the physical heavy-doublet eigenvalue, the two split octets, the complete messenger threshold, and these declared GUT-scale terms.

The measured low-energy couplings are evolved to $m_{\rm SUSY}$ in the $\overline{\rm MS}$ scheme and converted there according to $\alpha_{i,\overline{\rm DR}}^{-1}=\alpha_{i,\overline{\rm MS}}^{-1}-C_i(G)/(12\pi)$ before the supersymmetric running.  Superpartner mass splittings are represented by the common threshold $m_{\rm SUSY}=2$ TeV.  The quoted crossing is consequently a gauge-sector benchmark rather than a precision global unification fit.

\begin{figure*}[t]
\centering
\includegraphics[width=0.485\textwidth]{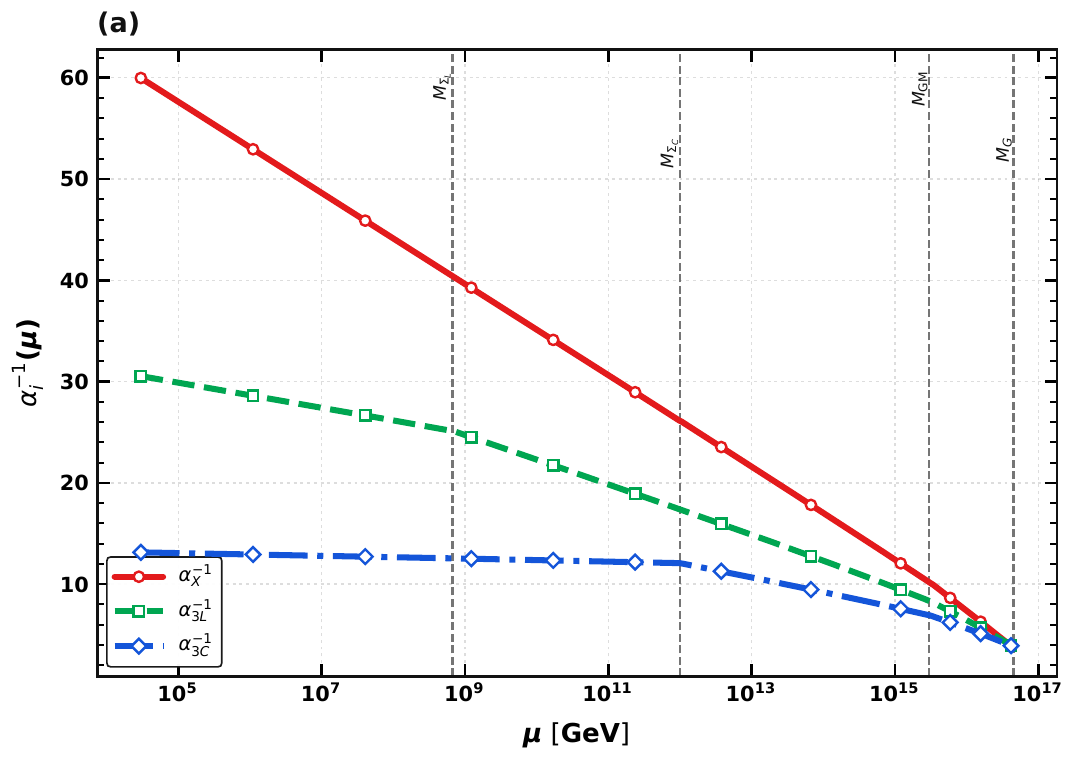}\hfill
\includegraphics[width=0.485\textwidth]{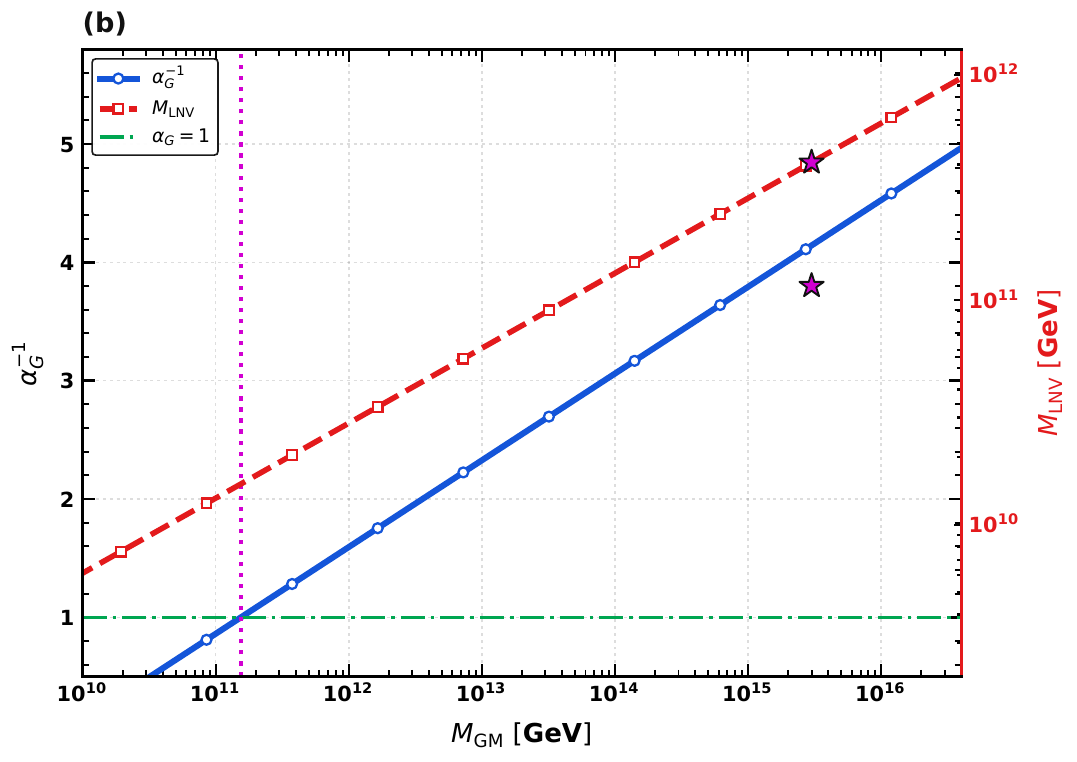}
\caption{Gauge-coupling evolution and the relation between the messenger and lepton-number scales.  Left: gauge-only two-loop evolution of the three $G_{331}$ gauge couplings, including the split-adjoint and complete-messenger thresholds.  Right: one-loop dependence of the unified inverse coupling and $M_{\rm LNV}$ on the charged-messenger threshold.  The star denotes $M_{\rm GM}=3.00\times10^{15}\,{\rm GeV}$, and the vertical dotted line corresponds to $\alpha_G=1$.}
\label{fig:gaugeconnection}
\end{figure*}

For two messenger pairs with $M_{\rm GM}=3.00\times10^{15}\,{\rm GeV}$, the two independent crossing conditions give
\begin{align}
M_{\Sigma_L}&=6.69\times10^8~{\rm GeV}, &
M_G&=4.49\times10^{16}~{\rm GeV},\nonumber\\
\alpha_G^{-1}&=3.80.&&
\label{eq:gaugeresult}
\end{align}
The corresponding loop parameter is $g_G^2/(16\pi^2)=2.09\times10^{-2}$.  Varying each conditional heavy-component mass ratio independently over $0.50$--$2.00$ gives $3.16\leq\alpha_G^{-1}\leq4.89$ and $2.05\times10^{16}\leq M_G/{\rm GeV}\leq6.65\times10^{16}$ for the 125-point scan.  This variation tests the sensitivity of the assumed completion; it does not replace the missing heavy mass matrices.  The evolution, including the weak-octet, color-octet, and messenger thresholds at their physical scales, is shown in \Fig{fig:gaugeconnection}(a).

We next use gauge mediation to determine the soft scale relevant to the $331$ vacuum.  At the messenger scale, the leading boundary conditions are~\cite{GiudiceRattazzi1999}
\begin{align}
M_a(M_{\rm GM})&=N_m\frac{\alpha_a(M_{\rm GM})}{4\pi}\Lambda_X,
\label{eq:gauginos}\\
m_i^2(M_{\rm GM})&=2N_m\Lambda_X^2
\sum_aC_a(i)\left[\frac{\alpha_a(M_{\rm GM})}{4\pi}\right]^2.
\label{eq:softmasses}
\end{align}
For the $\Phi$ triplet, the relevant Casimirs are $C_X=1/12$ and $C_{3L}=4/3$.  Evolving the gauge and gaugino contributions to $v_{331}$ gives
\begin{equation}
m_\Phi(v_{331})=7.76~{\rm TeV}.
\label{eq:mphisoft}
\end{equation}
After matching to the MSSM and evolving to $2$ TeV, the leading gaugino masses are
\begin{equation}
(M_1,M_2,M_3)=(0.873,1.70,4.93)~{\rm TeV}.
\label{eq:gauginolow}
\end{equation}
These running masses set the soft scale used in the $331$-breaking calculation.  The evolution of $m_\Phi$ retains only gauge and gaugino terms; contributions involving $Y$, $\lambda_H$, $A_\kappa$, $m_S^2$, and possible singlet tadpoles are not included.  Thus $m_\Phi$, $v_0$, and $\Delta_\Phi$ are leading benchmark estimates.  A precision electroweak spectrum also requires the third-generation Yukawa couplings and the electroweak symmetry-breaking conditions.

The benchmark has $F_X=\Lambda_XM_{\rm GM}=9.00\times10^{20}\,{\rm GeV}^2$, corresponding to $m_{3/2}=F_X/(\sqrt3\Mp)=213\,{\rm GeV}$.  Generic gravity-mediated terms are consequently not parametrically negligible compared with the lightest gauge-mediated gaugino mass.  The quoted gauge-mediated spectrum assumes sequestering or another suppression of flavor-dependent Planck-scale operators; without such a mechanism, the spectrum and flavor analysis then include gravity-mediated soft terms~\cite{GiudiceRattazzi1999}.

Equal soft masses for $\Phi$ and $\bar\Phi$ shift the radial minimum.  Along the direction in \Eq{eq:dflat}, the vacuum potential is
\begin{equation}
V(\varphi)=\frac{\kappa^2}{16}(\varphi^2-2v_0^2)^2
+\frac12m_\Phi^2\varphi^2.
\label{eq:radialpotential}
\end{equation}
Imposing $\partial_\varphi V=0$ at $\varphi=\sqrt2v_{331}$ gives
\begin{equation}
v_0^2=v_{331}^2+\frac{2m_\Phi^2}{\kappa^2}.
\label{eq:v0shift}
\end{equation}
The second derivative at the physical minimum contains no explicit soft-mass contribution,
\begin{equation}
m_\rho^2\equiv
\left.\frac{\partial^2V}{\partial\varphi^2}\right|_{\sqrt2v_{331}}
=\kappa^2v_{331}^2.
\label{eq:radialmass}
\end{equation}
The cancellation in \Eq{eq:radialmass} follows from the stationarity condition and holds for arbitrary $m_\Phi/v_{331}$.  Once $v_{331}$ is specified, the CMB normalization of $\kappa$ in \Sec{sec:inflation} determines the physical radial-mode mass.  Combining \Eqs{eq:v0shift}{eq:radialmass} gives the dimensionless vacuum-sensitivity ratio
\begin{equation}
\Delta_\Phi\equiv
\frac{2m_\Phi^2}{\kappa^2v_{331}^2}
=2\left(\frac{m_\Phi}{m_\rho}\right)^2
=\frac{v_0^2}{v_{331}^2}-1.
\label{eq:softradialratio}
\end{equation}
This ratio measures the hierarchy required when the CMB normalization and the gauge-mediated soft mass are imposed simultaneously.

Figure~\ref{fig:gaugeconnection}(a) shows the gauge-only two-loop RG evolution with the adopted weak-octet, color-octet, and messenger thresholds.  The three inverse couplings meet at the calculated conditional scale $M_G$, but the location of this crossing depends on the assumed heavy spectrum, including the unlifted adjoint states identified above. Figure~\ref{fig:gaugeconnection}(b) displays the one-loop messenger dependence of $\alpha_G^{-1}$ and the mediator mass $M_{\rm LNV}$.  The point marked by a star uses $M_{\rm GM}=3.00\times10^{15}$ GeV; the $\alpha_G=1$ reference line separates the perturbative domain in this simplified criterion. The two curves therefore illustrate a conditional relationship, not an independently predicted supersymmetric mass spectrum.

\section{Proton decay}\label{sec:proton}
The unification solution determines the dimension-six proton-decay contribution mediated by the broken unified gauge bosons.  For $p\to e^+\pi^0$, the decay width is~\cite{Weinberg1979,Wilczek1979,NathPerez2007,Hisano2022}
\begin{equation}
\Gamma_6=\frac{m_p}{64\pi f_\pi^2}
\frac{g_G^4}{M_X^4}A_R^2|\alpha_H|^2
(1+D+F)^2(1+|V_{ud}|^2)^2.
\label{eq:protonwidth}
\end{equation}
Here $M_X$ is the physical mass of the baryon-number-violating gauge boson, $A_R$ is the renormalization factor between the unified and hadronic scales, and $\alpha_H$, $D$, and $F$ are the hadronic matrix element and chiral parameters.  The physical vector mass is obtained from the adjoint kinetic term and need not coincide with the gauge-coupling crossing scale.  For the vacuum in \Eq{eq:avev}, a canonically normalized broken generator connecting the two $3\times3$ blocks satisfies
\begin{equation}
\Tr\!\left([T,\langle A\rangle]^\dagger[T,\langle A\rangle]\right)=\frac{v_A^2}{6},
\qquad
M_X=\frac{g_Gv_A}{\sqrt3}.
\label{eq:xbosonmass}
\end{equation}
For the calculated unified coupling, the choice $M_X=M_G$ corresponds to $v_A=4.28\times10^{16}\,{\rm GeV}$ and gives
\begin{equation}
\tau(p\to e^+\pi^0)=1.84\times10^{36}~{\rm yr}.
\label{eq:tau6}
\end{equation}
The Super-Kamiokande bound $\tau/B>2.4\times10^{34}\,{\rm yr}$~\cite{SuperK2020} implies
\begin{equation}
M_X>1.52\times10^{16}~{\rm GeV},
\qquad
v_A>1.44\times10^{16}~{\rm GeV},
\label{eq:mxbound}
\end{equation}
These bounds use the other inputs specified in \Eq{eq:protonwidth}.  The lifetime scales as $M_X^4$, as follows from \Eq{eq:protonwidth}.

Colored Higgsinos also generate dimension-five baryon-number-violating operators.  Their coefficients depend on the inverse colored-triplet mass matrix projected onto the triplet states coupled by the Yukawa interactions,
\begin{equation}
C_{5L}^{ijkl}\propto
(Y_u)^{ij}\,\mathcal P_T\,(Y_d)^{kl},
\qquad
\mathcal P_T\equiv(M_T^{-1})_{ab},
\label{eq:c5effective}
\end{equation}
An analogous coefficient is obtained for the RRRR operator.  Wino and higgsino dressing converts \Eq{eq:c5effective} into four-fermion operators.  The corresponding rate is proportional to $|\mathcal P_T|^2$, together with the Yukawa rotations and superpartner dressing functions~\cite{SakaiYanagida1982,Weinberg1982,DimopoulosRabyWilczek1982,HisanoMurayamaYanagida1993,MurayamaPierce2002,NathPerez2007,Hisano2022}.  The Super-Kamiokande limit $\tau/B(p\to K^+\bar\nu)>5.9\times10^{33}\,{\rm yr}$~\cite{SuperK2014Knu} consequently constrains this combination rather than a universal triplet mass.  For a multi-TeV superpartner spectrum, the dressing functions can suppress the rate sufficiently for GUT-scale triplets, whereas lighter sfermions or unsuppressed flavor projections require a smaller inverse propagator~\cite{Hisano2022,SuperK2014Knu,HyperK2018}.  The two-doublet condition in \Eq{eq:doubletcondition} does not determine the colored-triplet mass matrix.  A numerical choice of $M_T$ consequently constitutes an additional Higgs-sector assumption.  We state the constraint in terms of $\mathcal P_T$.  A numerical dimension-five lifetime can be calculated once a doublet--triplet completion specifies the triplet mass matrix and the Yukawa projections.

\section{Inflation from the 331-breaking fields}\label{sec:inflation}
The $\Phi+\bar\Phi$ pair responsible for breaking $G_{331}$ also defines the real $D$-flat direction in \Eq{eq:dflat}.  Nonminimally coupled scalar inflation has been studied in nonsupersymmetric theories and in supergravity~\cite{FutamaseMaeda1989,FakirUnruh1990,Kaiser1995,KomatsuFutamase1999,BezrukovShaposhnikov2008,BarbonEspinosa2009,BurgessLeeTrott2009,LernerMcDonald2010,KalloshLinde2010,BezrukovMagnin2011,Moursy2021}.  In the present model, inflation is controlled by the coupling $\kappa$, which also enters the $331$ vacuum potential in \Eq{eq:radialpotential}.  We take
\begin{multline}
K=-3\Mp^2\ln\!\Bigg[1-
\frac{|\Phi|^2+|\bar\Phi|^2+|S|^2}{3\Mp^2}
+\frac{\chi(\bar\Phi\Phi+{\rm h.c.})}{2\Mp^2}\\
+\frac{\zeta_S|S|^4}{3\Mp^4}\Bigg],
\label{eq:kahlerinflation}
\end{multline}
We use this K\"ahler potential together with the $\kappa$ term in \Eq{eq:superpotential}.  Along \Eq{eq:dflat}, we define $\xi\equiv\chi/4-1/6$.  The coefficient $\xi$ is an independent K\"ahler parameter.  We use $\xi=100$ as a representative point on the large-$\xi$ slow-roll plateau; this choice is independent of the particle spectrum.  The conformal factor is $\Omega^2=1+\xi\varphi^2/\Mp^2$, and the Einstein-frame potential becomes
\begin{equation}
V_E(\varphi)=
\frac{\kappa^2(\varphi^2-2v_0^2)^2}
{16(1+\xi\varphi^2/\Mp^2)^2}.
\label{eq:ve}
\end{equation}
The canonical field $\widehat\varphi$ satisfies
\begin{equation}
\left(\frac{d\widehat\varphi}{d\varphi}\right)^2=
\frac{1+\xi(1+6\xi)\varphi^2/\Mp^2}
{(1+\xi\varphi^2/\Mp^2)^2}.
\label{eq:metric}
\end{equation}
The mass parameter $v_0$ is fixed by \Eq{eq:v0shift}.  Although $v_0/\varphi\ll1$ during observable inflation, we retain the complete expression in \Eq{eq:ve} in the numerical calculation.

We denote by $N_*$ the number of e-folds between horizon exit of the pivot CMB scale and the end of inflation.  A star indicates evaluation at horizon exit.  In terms of $x\equiv\xi\varphi^2/\Mp^2$, the e-fold integral is
\begin{equation}
N_*=
\frac{1+6\xi}{8\xi}(x_*-x_e)
-\frac34\ln\frac{1+x_*}{1+x_e},
\label{eq:efolds}
\end{equation}
Here the endpoint $x_e$ is defined by $\epsilon=1$.  Once $x_*$ is obtained, the scalar amplitude $A_s$ determines $\kappa$.  The inflationary observables are
\begin{equation}
n_s=1-6\epsilon_*+2\eta_*,
\qquad
r=16\epsilon_*,
\qquad
\alpha_s\equiv\frac{dn_s}{d\ln k}.
\label{eq:observables}
\end{equation}

The CMB normalization can also be expressed as a relation for the radial mass in the broken phase.  Defining $y_*\equiv\varphi_*/\Mp$ and neglecting terms of order $v_0^2/\varphi_*^2$ only in the following analytical expression, we have
\begin{equation}
A_s=\frac{\kappa^2y_*^4}
{384\pi^2\epsilon_*(1+\xi y_*^2)^2}.
\label{eq:askappa}
\end{equation}
Using $r_*=16\epsilon_*$ gives
\begin{align}
\kappa&=\sqrt{24\pi^2A_sr_*}\,
\frac{1+\xi y_*^2}{y_*^2},
\label{eq:kappacmbclosed}\\
m_\rho&=v_{331}\sqrt{24\pi^2A_sr_*}\,
\frac{1+\xi y_*^2}{y_*^2}.
\label{eq:mrhocmbclosed}
\end{align}
For a trajectory specified by $(\xi,N_*)$, the amplitude $A_s$ sets the normalization, and Eq.~\eqref{eq:mrhocmbclosed} determines the radial mass.  Combining \Eq{eq:kappacmbclosed} with \Eq{eq:softradialratio} gives
\begin{equation}
\Delta_\Phi=
\frac{m_\Phi^2y_*^4}
{12\pi^2A_sr_*v_{331}^2(1+\xi y_*^2)^2}.
\label{eq:deltacmbclosed}
\end{equation}
Equations~\eqref{eq:mrhocmbclosed} and \eqref{eq:deltacmbclosed} display separately the dependence on the observational normalization and on the remaining model parameters.

At $(\xi,N_*)=(100,55)$,
\begin{align}
n_s&=0.965, & r&=3.50\times10^{-3},\nonumber\\
\alpha_s&=-6.21\times10^{-4}, &
\kappa&=4.23\times10^{-3}.
\label{eq:inflationnumbers}
\end{align}
Through \Eq{eq:radialmass}, the scalar normalization then gives
\begin{equation}
m_\rho=\kappa v_{331}=127~{\rm GeV}.
\label{eq:mrhonumber}
\end{equation}
Using the gauge-mediated soft mass in \Eq{eq:mphisoft}, \Eq{eq:v0shift} gives $v_0=2.59\times10^6\,{\rm GeV}$.  \Eq{eq:softradialratio} then gives $\Delta_\Phi=7.48\times10^3$, corresponding to $v_0/v_{331}=86.5$.  Thus the simultaneous vacuum and CMB conditions require a hierarchy between the superpotential mass parameter and the physical $331$ scale.  The collider production and Higgs-signal effects of the radial state depend on its mixing with the electroweak doublets in the complete scalar mass matrix.  This mixing remains an independent parameter of the Higgs sector considered here.

\begin{figure*}[t]
\centering
\includegraphics[width=0.485\textwidth]{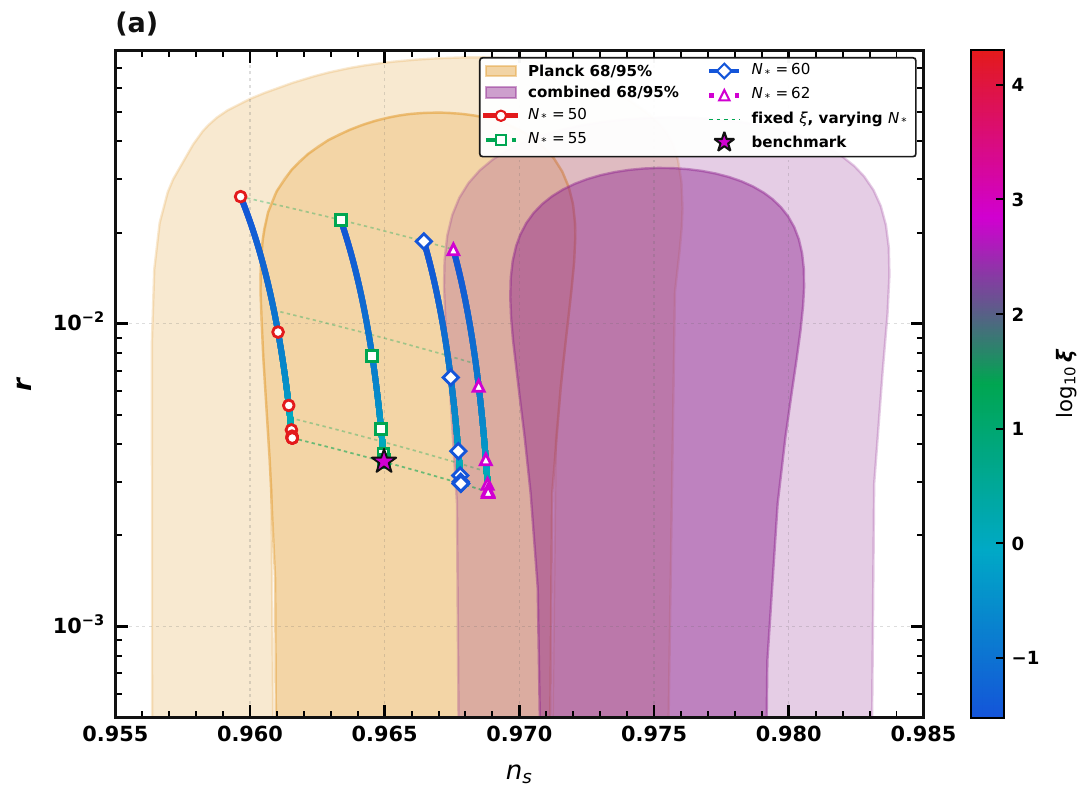}\hfill
\includegraphics[width=0.485\textwidth]{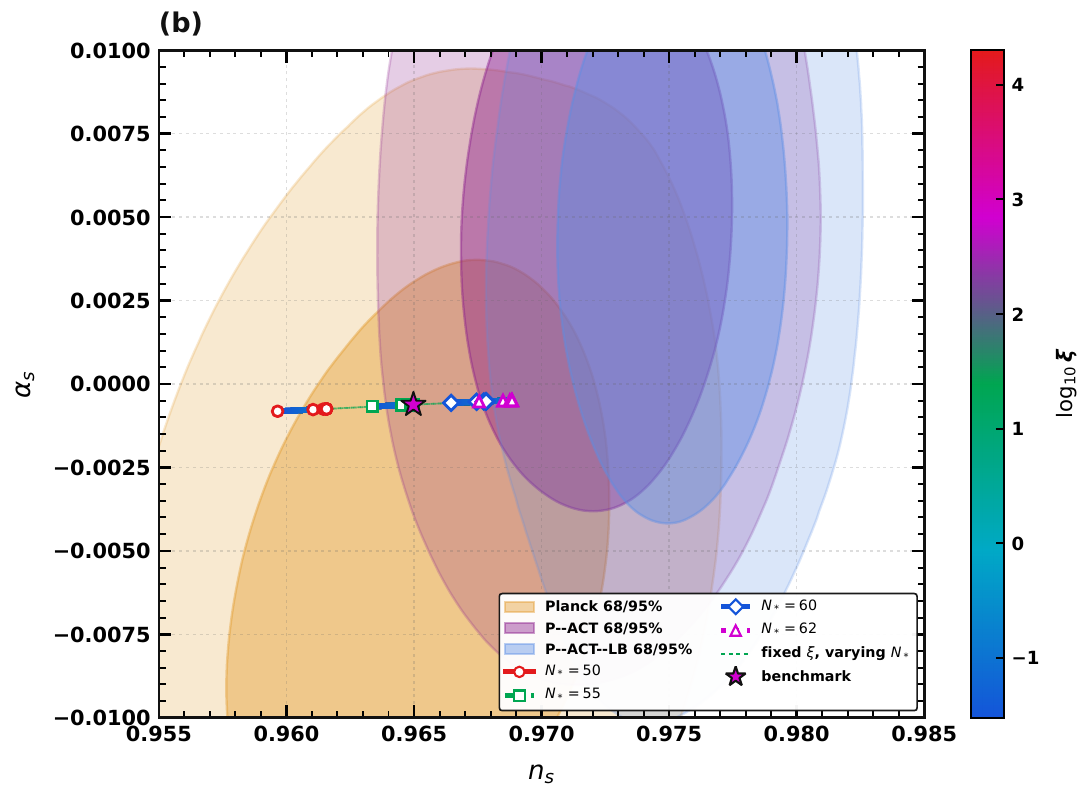}
\caption{Inflationary trajectories in the $(n_s,r)$ and $(n_s,\alpha_s)$ planes.  In the left panel $r$ denotes the tensor-to-scalar ratio at the CMB pivot scale. The shaded regions use the embedded contour coordinates for the cited Planck and ACT analyses. The color scale denotes $\xi$ along the fixed-$N_*$ trajectories; the transverse curves have fixed $\xi$ and varying $N_*$.  The right panel contains the Planck, P--ACT, and P--ACT--LB two-dimensional confidence regions.  The star denotes $(\xi,N_*)=(100,55)$.}
\label{fig:inflationobs}
\end{figure*}

In \Fig{fig:inflationobs}, the restricted inflationary trajectories are compared with the embedded observational contour polygons.  The $(n_s,\alpha_s)$ panel contains the Planck, P--ACT, and P--ACT--LB regions in the extended cosmological parameter space~\cite{PlanckInflation2018,ACTExtended2025}.  The $(n_s,r)$ panel uses combined CMB constraints that include BICEP/Keck~\cite{BICEP2021,Balkenhol2025}.  Along each fixed-$N_*$ curve, the color scale denotes $\xi$ over $0.03\le\xi\le2.00\times10^4$.  The line styles correspond to $N_*=50,55,60,62$, while the transverse curves have fixed representative values of $\xi$ and varying $N_*$.  The shaded boundaries are embedded numerical contour coordinates used for illustration, rather than a fresh posterior reconstruction from the CMB likelihoods. The label $r$ in Fig.~\ref{fig:inflationobs}(a) denotes the tensor-to-scalar ratio at the stated pivot. Future CMB polarization experiments will probe the same range of tensor amplitudes~\cite{CMBS4Science,LiteBIRD2023}.

\begin{figure}[t]
\centering
\includegraphics[width=0.98\columnwidth]{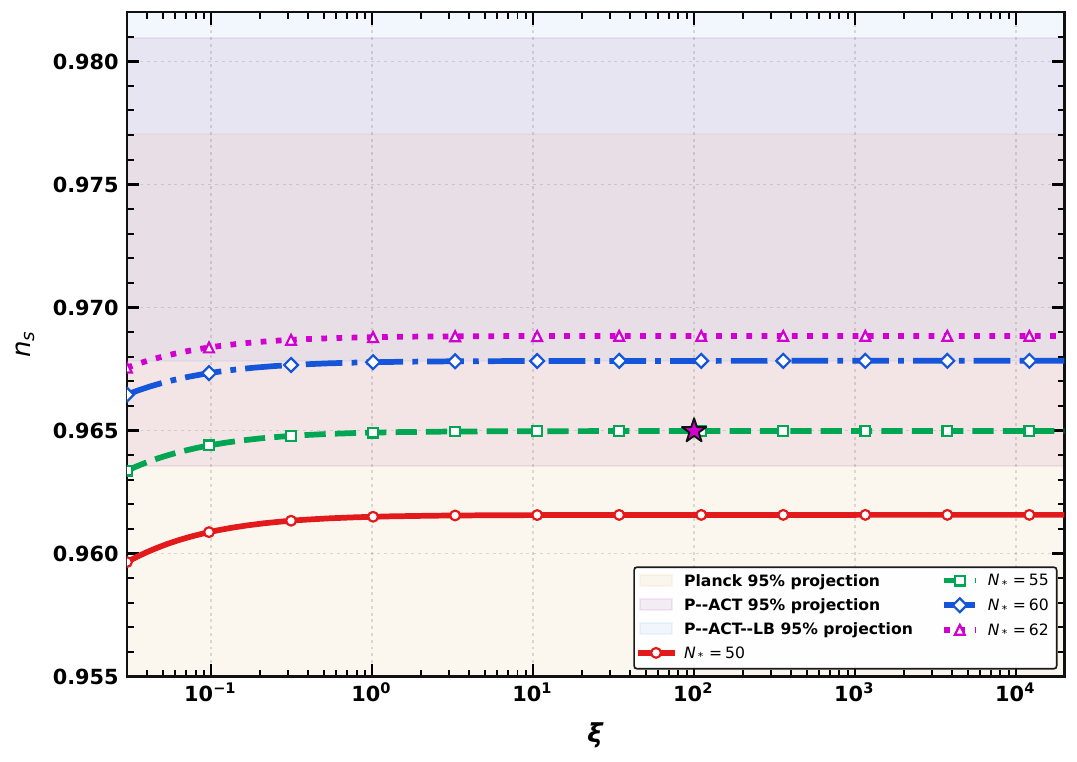}
\caption{$n_s$ as a function of $\xi$ for $N_*=50,55,60,62$.  The shaded intervals are geometric projections of the two-dimensional 95\% confidence regions in the $(n_s,\alpha_s)$ plane onto the $n_s$ axis; they are not marginalized one-dimensional intervals.  The star denotes $(\xi,N_*)=(100,55)$.}
\label{fig:nsxi}
\end{figure}

The two-dimensional confidence regions retain correlations between $n_s$ and the running that are absent from a comparison with $n_s$ alone.  Over the parameter range shown, the trajectory first enters the P--ACT 95\% region at $N_*\simeq53.3$ and the P--ACT--LB 95\% region at $N_*\simeq60.5$.  For fixed $\xi=100$, the latter intersection occurs at the same value within the quoted precision, and \Eq{eq:mrhocmbclosed} gives $m_\rho\simeq116~{\rm GeV}$.  Within the restricted potential, the contour-intersection condition maps to a corresponding radial mass; the interpretation depends on the assumed heavy-field stability and on the reheating history used to determine $N_*$~\cite{GarciaBellido2009}.  Figure~\ref{fig:nsxi} shows $n_s(\xi)$ for four fixed values of $N_*$.  The nearly horizontal curves at larger $\xi$ show that the e-fold number has a larger effect on the tilt than further increases of the nonminimal coupling in this regime. The shaded $n_s$ intervals are projections of the two-dimensional boundaries; they must not be interpreted as marginalized one-parameter bounds.

\begin{figure*}[t]
\centering
\includegraphics[width=0.485\textwidth]{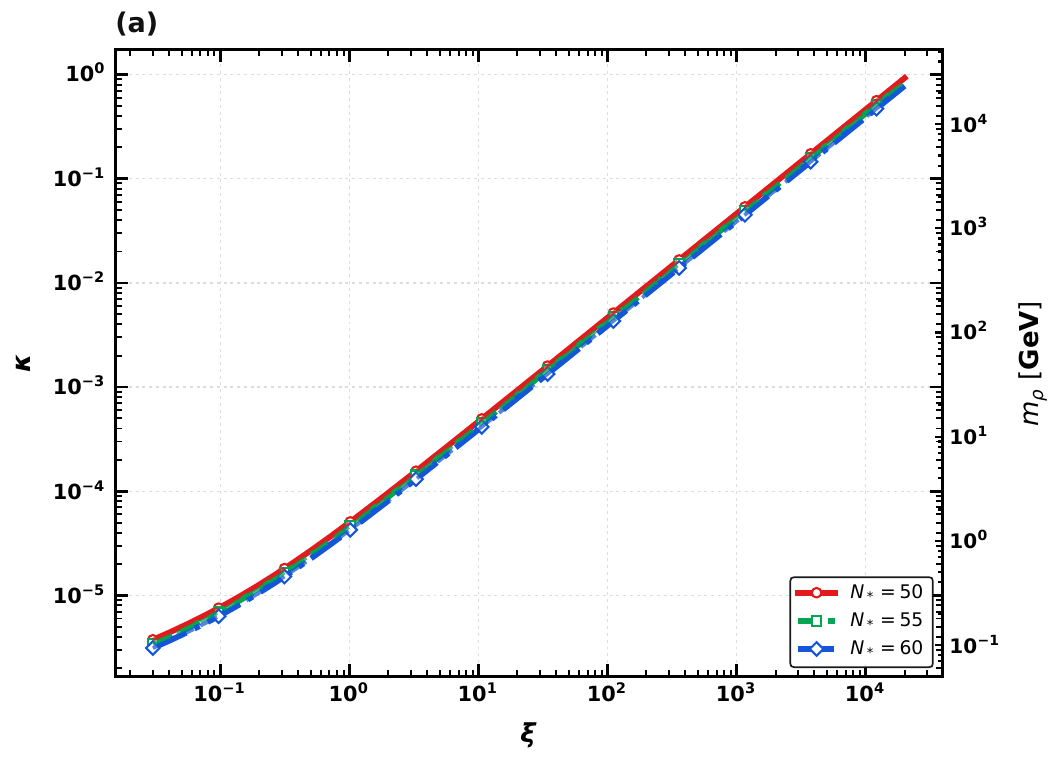}\hfill
\includegraphics[width=0.485\textwidth]{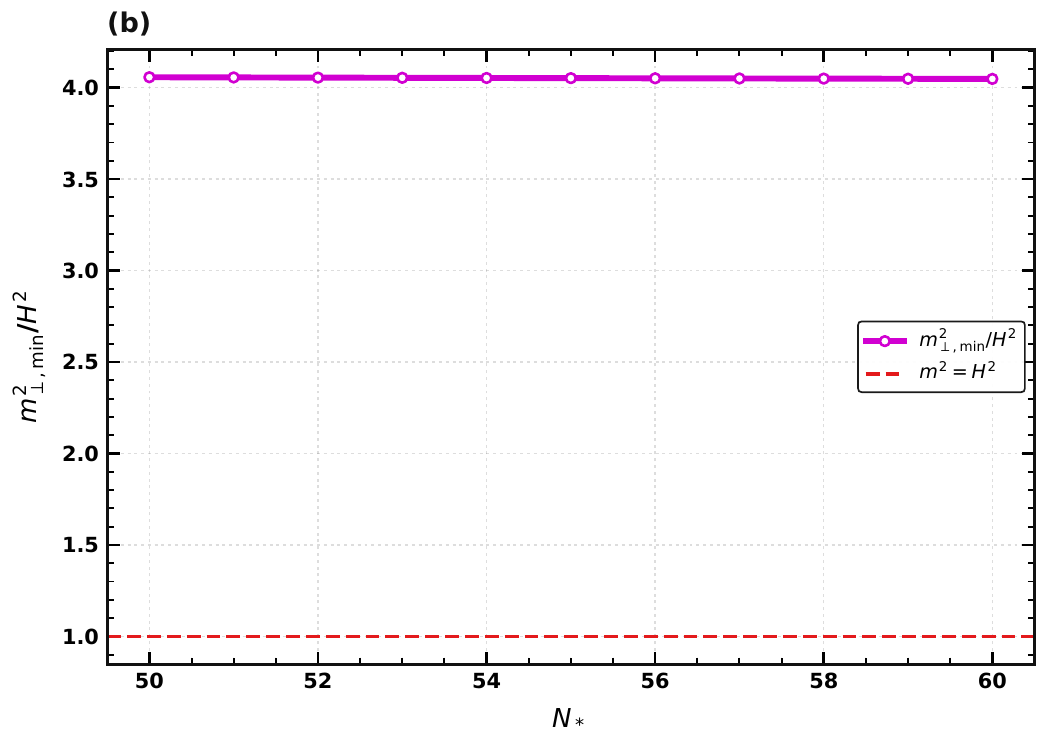}
\caption{Inflationary normalization and the $331$ radial-mode mass.  Left: $\kappa(\xi)$ fixed by $A_s$ and the corresponding mass $m_\rho=\kappa v_{331}$ for $N_*=50,55,60$, shown on the left and right vertical axes, respectively.  Right: smallest physical eigenvalue orthogonal to the inflaton in the covariant supergravity Hessian along the $\xi=100$ trajectory.}
\label{fig:inflationbridge}
\end{figure*}

We examine the stability of the trajectory in the neutral $(\Phi,\bar\Phi,S)$ supergravity subspace.  For the three complex fields $z^i=(\Phi,\bar\Phi,S)$, the F-term potential is
\begin{align}
V_F&=e^{K/\Mp^2}\left[
K^{i\bar j}D_iW D_{\bar j}\bar W
-\frac{3|W|^2}{\Mp^2}\right],\nonumber\\
D_iW&=\partial_iW+\frac{K_i}{\Mp^2}W.
\label{eq:sugraf}
\end{align}
For the neutral components, $T_{8L}(\Phi_3)=-1/\sqrt3$ and $X(\Phi_3)=1/(2\sqrt3)$.  Defining $P=K_\Phi\Phi-K_{\bar\Phi}\bar\Phi$, the broken diagonal generators give
\begin{equation}
V_D=\frac12\left(\frac{g_L^2}{3}+\frac{g_X^2}{12}\right)P^2.
\label{eq:dterm}
\end{equation}
Denoting by $G_{AB}$ the six-dimensional real-field metric obtained from $K_{i\bar j}$, the scalar mass matrix along the trajectory is determined by the generalized covariant Hessian
\begin{equation}
{\cal H}_{AB}=\nabla_A\nabla_BV,
\qquad
{\cal H}_{AB}u^B=m^2G_{AB}u^B.
\label{eq:covhessian}
\end{equation}
We evaluate the stability using the displayed K\"ahler potential and superpotential, without introducing an additional stabilizing term.  At horizon exit, for $\zeta_S=1$, the eigenvalues in units of $H_*^2$ are
\begin{equation}
(-5.06\times10^{-2},\;3.25\times10^{-3},\;4.05,\;946,\;946,\;4.19\times10^7).
\label{eq:hessianvalues}
\end{equation}
The negative eigenvalue is tangent to the inflaton trajectory and represents the local concavity in the rolling direction.  The small eigenvalue follows the broken gauge orbit and is removed after gauge fixing.  The four remaining scalar eigenvalues transverse to the inflaton are positive, with $m_\perp^2/H_*^2=4.05$ for the lightest mode.  This establishes stability only in the six-real-field neutral subspace.  It does not test the charged and colored components of $\Phi+\bar\Phi$, the adjoint fields, the Higgs sector, or the supersymmetry-breaking spurion.  Figure~\ref{fig:inflationbridge}(a) shows how the scalar-amplitude normalization fixes $\kappa$ and hence $m_\rho=\kappa v_{331}$ for the chosen $N_*$.  The right axis is an algebraic rescaling of the left axis, not an independent mass calculation. Figure~\ref{fig:inflationbridge}(b) displays the smallest physical transverse eigenvalue obtained from the neutral $(\Phi,\bar\Phi,S)$ Hessian for $50\le N_*\le60$. Its positivity establishes stability in that neutral sector only; it gives no mass bound on charged, colored, or adjoint fields.

The low-energy K\"ahler operator in \Eq{eq:kahleriss} applies after formation of the $331$ vacuum.  The inflationary potential in \Eq{eq:sugraf} is instead determined by the renormalizable $SU(6)$ superfields and the K\"ahler geometry in \Eq{eq:kahlerinflation}.  At the benchmark, $\varphi_*=2.15\times10^{18}\,{\rm GeV}$ and $H_*=1.47\times10^{13}\,{\rm GeV}$, while the weak-background estimate $\Mp/\xi=2.44\times10^{16}\,{\rm GeV}$ is close to the GUT scale.  Since $\varphi_*/M_G\simeq47.9$, control of higher-dimensional $SU(6)$ and K\"ahler operators and heavy-field backreaction is required in the ultraviolet completion.  The small ratio $H_*/M_G=3.27\times10^{-4}$ avoids direct excitation of GUT-scale modes but does not by itself establish effective-field-theory control~\cite{BarbonEspinosa2009,BurgessLeeTrott2009,BezrukovMagnin2011}.

\section{Analytical scale relations and parameter constraints}\label{sec:feasibility}
The calculation contains independent model parameters and quantities fixed by matching conditions or observational normalization.  These categories are listed separately in \Tab{tab:parameters}.  We first consider the relation induced by complete gauge-messenger multiplets.  At one loop, $N_m$ complete $\mathbf6+\overline{\mathbf6}$ pairs shift the three beta-function coefficients by the same amount.  The relative crossing scale is consequently unchanged, whereas
\begin{equation}
\alpha_G^{-1}=\alpha_{G,0}^{-1}
-\frac{N_m}{2\pi}\ln\frac{M_G}{M_{\rm GM}},
\label{eq:alphagmess}
\end{equation}
Here $\alpha_{G,0}$ is the unified coupling in the absence of the charged messengers.

The relation to the inverse-seesaw mediator is obtained at the same order.  Eliminating $M_{\rm GM}$ between \Eqs{eq:mlnv}{eq:alphagmess} gives
\begin{equation}
\alpha_G^{-1}(M_{\rm LNV})=
\alpha_{G,0}^{-1}-\frac{N_m}{2\pi}
\ln\!\left(\frac{M_G\Lambda_Xv_{331}^2}
{2\mu_0M_{\rm LNV}^3}\right),
\label{eq:alphagmlnv}
\end{equation}
It follows that
\begin{equation}
\frac{\partial\alpha_G^{-1}}{\partial\ln M_{\rm LNV}}
=\frac{3N_m}{2\pi}.
\label{eq:alphagmlnvslope}
\end{equation}
At one loop, this logarithmic slope is independent of the flavor coefficient $c_{ij}$ and the heavy-component threshold ratios.  These quantities affect the normalization and higher-order corrections.  For fixed $M_{\rm GM}$, the criterion $\alpha_G<1$ also gives an upper bound on the number of complete messenger pairs,
\begin{equation}
N_m<N_m^{\max}\equiv
\frac{2\pi(\alpha_{G,0}^{-1}-1)}{\ln(M_G/M_{\rm GM})}.
\label{eq:nmesscap}
\end{equation}
Without the messenger contribution, the conditional crossing occurs at $\alpha_{G,0}^{-1}=4.98$ and $M_G=4.09\times10^{16}\,{\rm GeV}$.  For $M_{\rm GM}=3.00\times10^{15}\,{\rm GeV}$, \Eq{eq:nmesscap} gives $N_m^{\max}=9.56$.  Thus the integer one-loop bound permits at most nine complete pairs for these thresholds, while the present model contains two.

\begin{figure}[t]
\centering
\includegraphics[width=0.98\columnwidth]{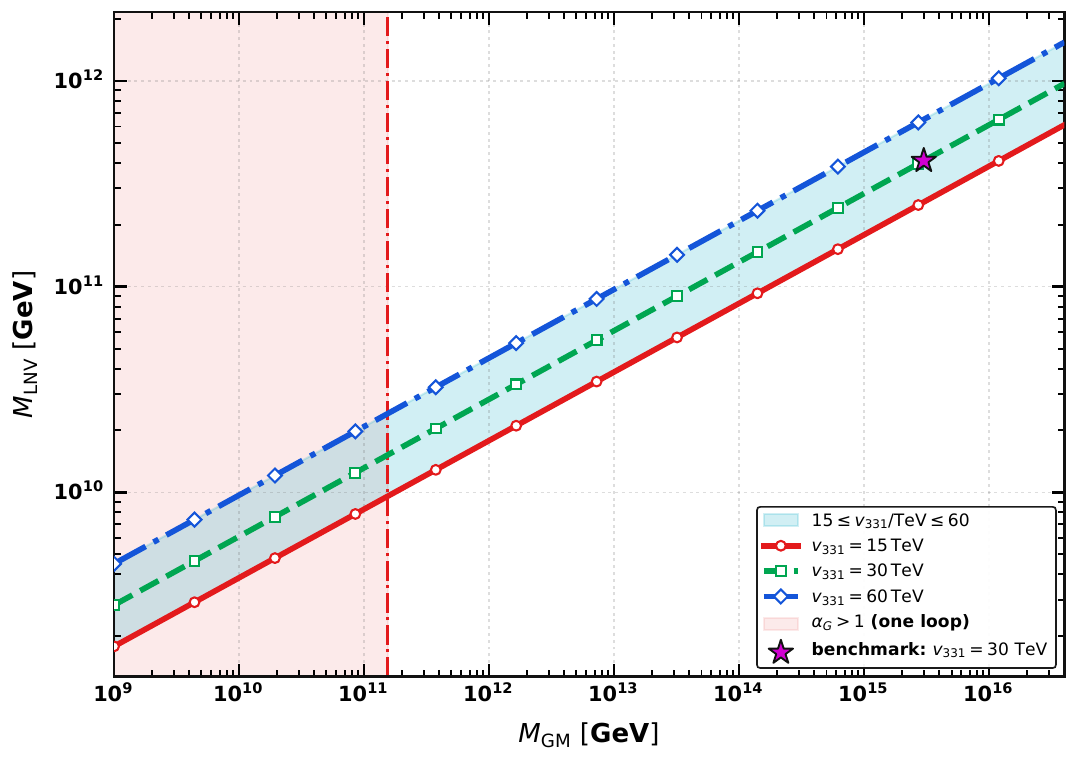}
\caption{Gauge-messenger and gauge-singlet mediator scales related by \Eq{eq:mlnv}.  The band corresponds to $15\le v_{331}/{\rm TeV}\le60$, and representative curves at fixed $v_{331}$ are shown.  The shaded region does not satisfy the one-loop condition $\alpha_G<1$.  The star denotes $v_{331}=30.0$ TeV and $M_{\rm GM}=3.00\times10^{15}\,{\rm GeV}$.}
\label{fig:spurionmap}
\end{figure}

Adopting the conventional criterion $\alpha_G<1$ yields
\begin{equation}
M_{\rm GM}>
M_G\exp\!\left[-\frac{2\pi}{N_m}(\alpha_{G,0}^{-1}-1)\right].
\label{eq:mgmbound}
\end{equation}
For $N_m=2$, the corresponding one-loop bound on the messenger scale is
\begin{equation}
M_{\rm GM}>2.82\times10^{10}~{\rm GeV}
\qquad(N_m=2,\ \alpha_G<1).
\label{eq:mgmboundnum}
\end{equation}
Combining \Eqs{eq:mlnv}{eq:mgmbound} gives the associated lower bound on the gauge-singlet lepton-number mediator scale,
\begin{equation}
M_{\rm LNV}>
\left[
\frac{\Lambda_Xv_{331}^2M_G}{2\mu_0}
\exp\!\left(-\frac{2\pi}{N_m}(\alpha_{G,0}^{-1}-1)\right)
\right]^{1/3}.
\label{eq:mlnvbound}
\end{equation}
For $v_{331}=30.0$ TeV, $\Lambda_X=3.00\times10^5\,{\rm GeV}$, and $M_{\rm GM}=3.00\times10^{15}\,{\rm GeV}$, we find $M_{\rm LNV}=4.09\times10^{11}\,{\rm GeV}$.  For the same low-energy neutrino normalization, \Eq{eq:mlnvbound} gives the lower limit $1.52\times10^{10}\,{\rm GeV}$.  The gauge-only two-loop solution for the assumed threshold spectrum also lies in this perturbative domain.  Figure~\ref{fig:spurionmap} shows the one-loop scaling of $M_{\rm LNV}$ with $M_{\rm GM}$ for $v_{331}=15$, $30$, and $60$ TeV.  The upper and lower curves delimit the displayed $v_{331}$ interval; the star marks the reference $30$ TeV benchmark. The lightly shaded nonperturbative region violates the illustrative condition $\alpha_G<1$.  The normalization of these curves assumes that lower-dimensional singlet-mass operators have been suppressed, as required by Sec.~\ref{sec:consistency}.

A second relation is obtained from the CMB normalization.  For a specified trajectory, Eqs.~\eqref{eq:kappacmbclosed} and \eqref{eq:mrhocmbclosed} determine $\kappa$ and $m_\rho$, while the gauge-mediated mass $m_\Phi$ shifts the physical vacuum according to \Eq{eq:v0shift}.  The same value of $\kappa$ is used in the inflationary and symmetry-breaking potentials.  At $(\xi,N_*)=(100,55)$, we find $\Delta_\Phi=7.48\times10^3$ and $v_0/v_{331}=86.5$.
The vacuum sensitivity may be expressed in the Barbieri--Giudice form~\cite{BarbieriGiudice1988,deCarlosCasas1993,AndersonCastano1995}.  From
$v_{331}^2=v_0^2-2m_\Phi^2/\kappa^2$, define
$\Delta_p\equiv|\partial\ln v_{331}^2/\partial\ln p|$.  Then
\begin{equation}
\Delta_{v_0^2}=1+\Delta_\Phi,
\qquad
\Delta_{m_\Phi^2}=\Delta_{\kappa^2}=\Delta_\Phi,
\label{eq:bgsensitivities}
\end{equation}
Here $\Delta_\Phi$ is defined in \Eq{eq:softradialratio}.  The value $\Delta_\Phi=7.48\times10^3$ corresponds to a cancellation of approximately $1.34\times10^{-4}$ in the minimal vacuum.  Multiplicative renormalization can preserve a small value of $\kappa$ against additive radiative corrections, but it does not remove the sensitivity in \Eq{eq:bgsensitivities}.

The third relation follows from the rank of the neutrino mass matrix.  The leading antisymmetric operator gives one vanishing Takagi mass.  Within the restricted displayed operator basis, the correction considered here has $\epsilon_\Phi\sim v_{331}^2/M_{\rm UV}^2$.  Since the inverse-seesaw determinant contains $[\det(m_D)]^2$, \Eq{eq:fourthsuppression} shows that the lightest mass is quadratic in the Dirac correction and fourth order in $v_{331}/M_{\rm UV}$.  This operator hierarchy gives a conditional scaling of the lightest eigenvalue; it does not establish a prediction for observed neutrino masses without a completed flavor/parity sector.

\begin{table}[t]
\centering
\caption{Parameter classification.  Experimental inputs and fitted flavor data are distinguished from independent model parameters, derived quantities, and constraints.}
\label{tab:parameters}
\scriptsize
\begin{tabular}{@{}l@{\hspace{0.6em}}l@{}}
\toprule
category & \parbox[t]{0.72\columnwidth}{quantities and role} \\
\midrule
external data & \parbox[t]{0.72\columnwidth}{SM gauge couplings, $A_s$, neutrino mass splittings, and mixing parameters; experimental matching and normalization inputs.} \\
model inputs & \parbox[t]{0.72\columnwidth}{$v_{331}$, $\Lambda_X$, $M_{\rm GM}$, $N_m$, $M_{\Sigma_C}$, $m_{\rm SUSY}$, $(\xi,N_*,\zeta_S)$, Higgs parameters, $M_N$, and GUT-fragment ratios.} \\
derived & \parbox[t]{0.72\columnwidth}{$\bar\lambda_H$, $\mu_0$, $M_{\rm LNV}$, $M_{\Sigma_L}$, $(M_G,\alpha_G)$, $\kappa$, $m_\rho$, $m_\Phi$, and $v_0$ from matching, reconstruction, RGE, CMB normalization, and vacuum equations.} \\
constraints & \parbox[t]{0.72\columnwidth}{$\alpha_G<1$, proton lifetime, positive physical orthogonal Hessian eigenvalues, and Planck/ACT confidence regions restrict the input space.} \\
\bottomrule
\end{tabular}
\end{table}

For $v_{331}=30.0$ TeV, $\Lambda_X=3.00\times10^5\,{\rm GeV}$, and $M_{\rm GM}=3.00\times10^{15}\,{\rm GeV}$, we obtain $M_{\rm LNV}=4.09\times10^{11}\,{\rm GeV}$, $M_G=4.49\times10^{16}\,{\rm GeV}$, $\alpha_G^{-1}=3.80$, $\kappa=4.23\times10^{-3}$, and $m_\rho=127$ GeV.  The unification values are conditional on the declared heavy spectrum, whereas the neutrino and inflationary formulas hold only within the specified effective subsectors.

In the conditional neutral-sector parametrization, the inverse-seesaw states lie at $2$--$3$ TeV and have $\max|\Theta_{\alpha i}|=2.33\times10^{-3}$.  The leading gauge-mediated gaugino masses at $2$ TeV are $(M_1,M_2,M_3)=(0.873,1.70,4.93)$ TeV.  The $331$ gauge-boson masses are of order $g_{331}v_{331}$, with coefficients determined by the corresponding broken generators.  These masses and mixings are the inputs for subsequent collider and flavor studies.

\section{Conclusions}\label{sec:conclusion}

We propose a $N=1$ supersymmetric $SU(6)$ theory with an intermediate $(SU(3)_C\times SU(3)_L\times U(1)_X)$ phase, in which common fields and operators relate the neutrinos, gauge mediation, unification, and symmetry-breaking sectors. The analysis identifies three analytical relations arising from the neutrino Yukawa structure, the common supersymmetry-breaking spurion, and the scalar sector
responsible for $331$ symmetry breaking. We have also examined the operator selection rules and heavy-field spectrum required for these relations to describe a consistent unified theory.

In the neutrino sector, the antisymmetric Dirac mass matrix has rank two, leading to one vanishing light-neutrino mass at leading order within the inverse-seesaw framework. We have shown that the leading rank-breaking correction is determined by its projection onto the left and right null vectors of the antisymmetric matrix. For a correction suppressed
by $\epsilon_\Phi\sim v_{331}^2/M_{\rm UV}^2$,
the lightest neutrino mass scales as
$m_{\rm lightest}\propto(v_{331}/M_{\rm UV})^4$,
provided that the corresponding flavor projection is nonzero. This result establishes the dependence of the smallest eigenvalue on the symmetry-breaking scale and the higher-dimensional flavor structure. However, the unbroken neutrino parity prevents the
ordinary active leptons from mixing with the
inverse-seesaw states in the displayed field assignment. Consequently, the illustrative neutrino reconstruction requires an additional flavor and parity-breaking completion before it can be interpreted as a description
of the observed neutrino spectrum.

In the gauge-mediation and unification sectors,
complete vector-like messenger multiplets preserve the relative one-loop gauge coupling evolution while modifying the unified coupling. When the same supersymmetry-breaking spurion enters the messenger and lepton-number-violating operators, their common
dependence leads to the analytical relation
$\partial\alpha_G^{-1}/\partial\ln M_{\rm LNV}
=3N_m/(2\pi)$. The coefficient is independent of the flavor normalization and the assumed heavy-threshold ratios at one loop. The gauge-only two-loop analysis yields
a perturbative unified coupling for the adopted
heavy-field spectrum, with the crossing scale above the conventional supersymmetric unification scale. The corresponding unified gauge-boson masses determine
the dimension-six proton-decay contribution, whereas the dimension-five contribution depends on the colored-triplet mass matrix and its flavor structure. The numerical unification and proton-decay results therefore remain dependent on the specified heavy-sector
completion.

In the inflationary sector, the fields responsible
for $331$ symmetry breaking also define a
nonminimally coupled inflationary trajectory.
The observed scalar amplitude determines the
superpotential coupling $\kappa$, which fixes
the radial mass through $m_\rho=\kappa v_{331}$
within the restricted scalar potential. This
establishes a relation between the inflationary
normalization and the symmetry-breaking spectrum.
The resulting inflationary observables can be
compared with current CMB constraints, while the
covariant supergravity mass matrix gives positive
physical eigenvalues transverse to the inflaton
within the neutral $(\Phi,\bar\Phi,S)$ subspace.
Combining the CMB-normalized coupling with the
gauge-mediated scalar mass, however, requires
a substantial cancellation in the symmetry-breaking
vacuum, corresponding to a sensitivity at the
level of $10^{-4}$ for the representative benchmark.

The simultaneous realization of these mechanisms
requires additional ultraviolet structure. In
particular, the stated symmetries permit direct
singlet Majorana masses and lower-dimensional
K\"ahler operators that can dominate the proposed lepton-number-violating contribution. Additional superpotential interactions can also modify the symmetry-breaking vacuum, while uneaten adjoint bifundamentals and colored Higgs states require a complete mass spectrum. The stability of the charged, colored, and other heavy scalar directions
during inflation must likewise be established.
These conditions distinguish the analytical
relations obtained within the specified effective subsectors from predictions of a fully completed $SU(6)$ theory. The results provide a framework for relating neutrino mass generation, supersymmetry breaking, gauge unification, and inflation, subject to a consistent realization of the required flavor, Higgs, and heavy-field sectors.

\begin{acknowledgments}

	TL is supported in part by the National Key Research and Development Program of China Grant No. 2020YFC2201504, by the Projects No. 11875062, No. 11947302, No. 12047503, and No. 12275333 supported by the National Natural Science Foundation of China, by the Key Research Program of the Chinese Academy of Sciences, Grant No. XDPB15, by the Scientific Instrument Developing Project of the Chinese Academy of Sciences, Grant No. YJKYYQ20190049, by the International Partnership Program of Chinese Academy of Sciences for Grand Challenges, Grant No. 112311KYSB20210012, and by the Henan Province Outstanding Foreign Scientist Studio Project, No.GZS2025008. The authors also acknowledge that ChatGPT was used solely for language editing during the preparation of this manuscript, while the authors take full responsibility for the scientific content and final manuscript.

\end{acknowledgments}

\appendix

\section{Representation and symmetry checks}\label{app:group}
The decomposition in \Eq{eq:branch15} follows directly from the exterior product of the two terms in \Eq{eq:branch6}.  For the color--color part, $\wedge^2\mathbf3=\overline{\mathbf3}$ and the $X$ charge adds to $-2q_6$.  The left--left part gives $\wedge^2\mathbf3_L=\overline{\mathbf3}_L$ with charge $+2q_6$, while the mixed product is $(\mathbf3,\mathbf3,0)$.  The conjugate antifundamental branches as
\begin{equation}
\overline{\mathbf6}=(\overline{\mathbf3},\mathbf1,+q_6)
\oplus(\mathbf1,\overline{\mathbf3},-q_6).
\label{eq:branch6barapp}
\end{equation}
These signs reproduce the usual electric charges through \Eq{eq:charge}.  The gauge beta coefficients depend on the Abelian charges through $X^2$, so these signs leave them unchanged.

For the chiral family, the cubic anomaly coefficient of the two-index antisymmetric representation of $SU(N)$ is $N-4$ in units with fundamental coefficient one.  At $N=6$, this gives $A(\mathbf{15})=2$, and leads to \Eq{eq:anomaly}.  Vectorlike pairs have zero net cubic anomaly.

The $Z_4^R$ assignments in \Eq{eq:rcharges} give charge two to all displayed renormalizable superpotential terms.  The $Z_2^\nu$ parity distinguishes $\bar6'$ from the second antifundamental and permits $\bar6'\Phi N_s$ while forbidding the corresponding mixing with the ordinary antifundamental.  For the adjoint shaping symmetry, $q_{Z_3^G}(A)=1$ and the light matter and Higgs multiplets have zero $Z_3^G$ charge.  Consequently
\begin{equation}
q(\bar6'\bar6'H_{15}A^n)=n\pmod3,
\label{eq:z3operatorcharge}
\end{equation}
so one- and two-adjoint insertions are forbidden while $n=3$ is allowed.  On the vacuum in \Eq{eq:avev}, $A^3$ is proportional to $A$, and the action of any polynomial in $A$ on the unbroken $SU(3)_L$ block is proportional to the identity.  These contractions retain the leading antisymmetry for the light neutral components.  The lower-order rank-breaking structure is the five-superfield operator in \Eq{eq:phirankoperator}.  Its total $Z_4^R$ charge is $1+0+2+1+2=2$ modulo four, its two $Z_2^\nu$-odd fields give even parity, and all fields in the operator are neutral under $Z_3^G$.  Hence the symmetries of the model permit it with coefficient $M_{\rm UV}^{-2}$.

\section{Gauge coefficients, messenger threshold, and soft running}\label{app:gauge}
For a chiral multiplet in $(R_C,R_L,X)$, the Abelian index and the two non-Abelian indices used in \Eq{eq:bone} are
\begin{align}
T_X&=X^2d(R_C)d(R_L),\nonumber\\
T_L&=T(R_L)d(R_C),\qquad
T_C=T(R_C)d(R_L).
\label{eq:dynkinproducts}
\end{align}
The corresponding Casimirs entering \Eq{eq:btwo} are
\begin{equation}
C_X=X^2,\qquad C_L=C(R_L),\qquad C_C=C(R_C).
\label{eq:casimirs}
\end{equation}
For the $SU(3)$ representations used here,
\begin{align}
T(\mathbf3)=T(\overline{\mathbf3})&=\frac12,\nonumber\\
C(\mathbf3)=C(\overline{\mathbf3})&=\frac43,\qquad
T(\mathbf8)=C(\mathbf8)=3.
\label{eq:su3indices}
\end{align}
Substitution of the field content in \Tab{tab:fields} reproduces \Eq{eq:coefficients}.  The matrix follows directly from the Dynkin indices, spectator multiplicities, and quadratic Casimirs in \Eqs{eq:bone}{eq:btwo}.

A complete messenger pair decomposes as
\begin{align}
\mathbf6+\overline{\mathbf6}={}&(\mathbf3,\mathbf1,-q_6)
+(\overline{\mathbf3},\mathbf1,+q_6)\nonumber\\
&+(\mathbf1,\mathbf3,+q_6)
+(\mathbf1,\overline{\mathbf3},-q_6).
\label{eq:messdecomp}
\end{align}
Using \Eq{eq:dynkinproducts}, the four terms give the one-loop sum
\begin{equation}
\Delta b^{(\mathbf6+\overline{\mathbf6})}=(1,1,1),
\label{eq:messbapp}
\end{equation}
as stated in \Eq{eq:messengerb}.  The corresponding gauge-only two-loop increment for one pair is
\begin{equation}
\Delta B_{ab}^{\rm mess}=
\begin{pmatrix}
1/3&8/3&8/3\\
1/3&34/3&0\\
1/3&0&34/3
\end{pmatrix}.
\label{eq:messB2}
\end{equation}
Two pairs multiply \Eqs{eq:messbapp}{eq:messB2} by two above $M_{\rm GM}$.

The one-loop analytical relation \Eq{eq:alphagmess} follows by integrating
\begin{equation}
\frac{d\alpha_i^{-1}}{d\ln\mu}=-\frac{b_i}{2\pi}
\label{eq:onealpharg}
\end{equation}
from $M_{\rm GM}$ to $M_G$.  A common $\Delta b=N_m$ shifts every inverse coupling equally and leaves the one-loop differences $\alpha_i^{-1}-\alpha_j^{-1}$ invariant.  The two-loop calculation used for \Eq{eq:gaugeresult} includes the nonuniversal matrix contribution in \Eq{eq:messB2}.

For one electroweak chiral doublet with $|Y|=1/2$, the GUT-normalized Abelian index is $(3/5)Y^2d(\mathbf2)=3/10$.  This is the contribution of one chiral doublet, whereas a vectorlike pair contains two such superfields.  The complete $N=1$ supersymmetric threshold is therefore
\begin{equation}
\Delta b_{\rm SM}^{D_H}=(3/5,1,0),
\qquad
\Delta b_{331}^{D_H}=(1/2,1,0),
\label{eq:doubletthreshold}
\end{equation}
where the second result follows from $\alpha_X^{-1}=(5\alpha_1^{-1}-\alpha_{3L}^{-1})/4$.  Integrating $d\alpha_i^{-1}/d\ln\mu=-b_i/(2\pi)$ across a state omitted below its mass gives the sign in \Eq{eq:finitematching}.  The vectors in \Eq{eq:heavyvectors} are reconstructed from \Eq{eq:dynkinproducts}; in particular, the color pair in a $\mathbf6+\overline{\mathbf6}$ has $2T_X=1/2$ and $2T_C=1$.

The leading $\Phi$ soft mass at $M_{\rm GM}$ follows from \Eq{eq:softmasses} with
\begin{equation}
C_X(\Phi)=q_6^2=\frac1{12},
\qquad C_{3L}(\Phi)=\frac43,
\qquad C_{3C}(\Phi)=0.
\label{eq:phicasimirs}
\end{equation}
Below the messenger threshold, the gauge contribution to the soft running is
\begin{align}
16\pi^2\frac{dm_\Phi^2}{d\ln\mu}
&=-8\sum_aC_a(\Phi)g_a^2|M_a|^2,\nonumber\\
16\pi^2\frac{dM_a}{d\ln\mu}&=2b_ag_a^2M_a.
\label{eq:softRGE}
\end{align}
We integrate \Eq{eq:softRGE} through the color- and weak-octet thresholds.  At $v_{331}$ the $U(1)_Y$ gaugino matches using the normalized generator employed in the gauge-coupling matching,
\begin{equation}
\frac{M_1}{\alpha_1}=\frac45\frac{M_X}{\alpha_X}
+\frac15\frac{M_{3L}}{\alpha_{3L}}.
\label{eq:gauginomatch}
\end{equation}
The values quoted in \Eq{eq:gauginolow} then follow from one-loop $M_a/\alpha_a$ invariance in the MSSM interval.

\section{Inverse-seesaw reconstruction and determinant expansion}\label{app:inverse}
Let $U$ denote the PMNS matrix and choose normal ordering with $m_1=0$.  The null vector of the target light-neutrino matrix is the first column of $U$.  For a complex vector $n=(n_1,n_2,n_3)^T$, the antisymmetric matrix
\begin{equation}
A(n)=\begin{pmatrix}
0&n_3&-n_2\\
-n_3&0&n_1\\
n_2&-n_1&0
\end{pmatrix}
\label{eq:Across}
\end{equation}
satisfies $A(n)n=0$.  We normalize $A$ by its Frobenius norm and set $m_D=m_D^{(0)}A$.  Defining
\begin{equation}
F\equiv m_DM_N^{-T},
\label{eq:Fmatrix}
\end{equation}
\Eq{eq:lightmass} becomes $m_\nu=F\mu_SF^T$.  On the rank-two image of $F$, a symmetric solution is
\begin{equation}
\mu_S^{\rm im}=F^+m_\nu(F^T)^+,
\label{eq:mupinv}
\end{equation}
Here $+$ denotes the Moore--Penrose inverse.  Components of $\mu_S$ along the null direction of $F$ lie outside the rank-two image and leave the leading light matrix invariant.  The neutrino reconstruction used here includes one such component to keep the singlet-sector coefficient matrix nonsingular; the rank argument depends only on the image component.

The determinant scaling in \Eq{eq:detmdexp} follows from the general first-order identity
\begin{equation}
\det(A+\epsilon B)=\det A+
\epsilon\Tr[\operatorname{adj}(A)B]+\mathcal O(\epsilon^2).
\label{eq:detidentity}
\end{equation}
For $\operatorname{rank}A=2$ in three dimensions, $\operatorname{adj}(A)$ has rank one.  If $r$ and $\ell$ are right and left null vectors, respectively, then
\begin{equation}
\operatorname{adj}(A)=\mathcal C\,r\ell^T
\label{eq:adjrankone}
\end{equation}
for a nonzero coefficient $\mathcal C$.  Inserting \Eq{eq:adjrankone} into \Eq{eq:detidentity} gives $\det m_D=\epsilon_\Phi\mathcal C\,\ell^T\delta m_Dr+\mathcal O(\epsilon_\Phi^2)$, giving \Eq{eq:detmdnull}.  Combining this expression with \Eq{eq:detderivation} and $|\det m_\nu|=m_{\rm lightest}m_2m_3$ yields the coefficient in \Eq{eq:mlightcoefficient}.

Taking the determinant of \Eq{eq:lightmass} assumes nonsingular $M_N$ and $\mu_S$.  Under these conditions,
\begin{align}
\det m_\nu
&=\det(m_D)\det(M_N^{-T})\det(\mu_S)
\det(M_N^{-1})\det(m_D^T)\nonumber\\
&=\frac{[\det(m_D)]^2\det(\mu_S)}{[\det(M_N)]^2},
\label{eq:detderivation}
\end{align}
giving \Eq{eq:detmnu}.  Since a complex symmetric Majorana matrix admits a Takagi factorization, the product of its physical masses is $|\det m_\nu|$.

\section{Inflationary derivatives and covariant supergravity masses}\label{app:inflation}
Along \Eq{eq:dflat}, define $y=\varphi/\Mp$ and $x=\xi y^2$.  Neglecting terms suppressed by $v_0^2/\varphi^2$ only for the analytical slow-roll expressions, the potential has the quartic form
\begin{equation}
V_E\simeq\frac{\lambda\varphi^4}{4(1+\xi y^2)^2},
\qquad \lambda=\frac{\kappa^2}{4}.
\label{eq:quarticapprox}
\end{equation}
Using \Eq{eq:metric}, the potential slow-roll parameter is
\begin{equation}
\epsilon=\frac{8}{y^2[1+\xi(1+6\xi)y^2]}.
\label{eq:epsilonapp}
\end{equation}
Solving $\epsilon(y_e)=1$ gives
\begin{equation}
x_e=\frac{-1+\sqrt{1+32\xi(1+6\xi)}}{2(1+6\xi)}.
\label{eq:xe}
\end{equation}
The canonical e-fold integral $N=\int V/(\Mp^2V_{,\widehat\varphi})\,d\widehat\varphi$ gives \Eq{eq:efolds}.  Once $x_*$ is known, the scalar amplitude
\begin{equation}
A_s=\frac{V_*}{24\pi^2\Mp^4\epsilon_*}
\label{eq:As}
\end{equation}
sets $\kappa$.  Substituting \Eq{eq:quarticapprox} gives
\begin{equation}
\kappa^2=384\pi^2A_s\epsilon_*
\frac{(1+\xi y_*^2)^2}{y_*^4},
\label{eq:kappaderivationapp}
\end{equation}
Using $r_*=16\epsilon_*$ gives \Eq{eq:kappacmbclosed}.  Multiplication by $v_{331}$ gives \Eq{eq:mrhocmbclosed}, and inserting the result into \Eq{eq:softradialratio} gives \Eq{eq:deltacmbclosed}.  The complete potential in \Eq{eq:ve} is used when evaluating $H_*$.

For the multifield calculation, write the dimensionless complex fields as $z^i/\Mp$.  The logarithm in \Eq{eq:kahlerinflation} defines a real function $\Omega$, $K=-3\Mp^2\ln\Omega$.  In Planck units the K\"ahler metric can be evaluated as
\begin{equation}
K_{i\bar j}=
3\frac{\Omega_i\Omega_{\bar j}}{\Omega^2}
-3\frac{\Omega_{i\bar j}}{\Omega}.
\label{eq:kahlerderivative}
\end{equation}
Equation~\eqref{eq:kahlerderivative} determines the six-real-field metric and its Christoffel symbols entering \Eq{eq:covhessian}.  At $(\xi,N_*)=(100,55)$ the metric eigenvalues are
\begin{equation}
(0.0127,0.0127,0.0127,0.0127,7.52,7.52),
\label{eq:metricnumbers}
\end{equation}
so the kinetic matrix is positive definite.  The diagonal charges of the neutral components give the $D$-term coefficient in \Eq{eq:dterm}; omitting these generator eigenvalues would overestimate that contribution by a factor of three.  The generalized Hessian eigenvalues are listed in \Eq{eq:hessianvalues}.  The eigenvector associated with the small value aligns with the relative phase of $\Phi$ and $\bar\Phi$, the broken gauge direction.  Removing this gauge orbit and the tangent direction leaves the four physical orthogonal scalar masses used in \Fig{fig:inflationbridge}.  This construction contains only the six real components of $(\Phi,\bar\Phi,S)$ and is not a Hessian of the complete $SU(6)$ field space.

The $(n_s,\alpha_s)$ comparison uses the two-dimensional Planck, P--ACT, and P--ACT--LB confidence boundaries associated with the extended cosmological analyses~\cite{PlanckInflation2018,ACTExtended2025}.  The $(n_s,r)$ comparison uses two-dimensional combined CMB constraints including BICEP/Keck~\cite{BICEP2021,Balkenhol2025}.  We preserve the two-dimensional geometry and retain their two-dimensional geometry instead of reconstructing them from separate one-dimensional errors.

\section{Proton normalization and the colored-triplet coefficient}\label{app:proton}
The dimension-six lifetime follows by inverting \Eq{eq:protonwidth}.  The numerical calculation uses
\begin{align}
m_p&=0.938~{\rm GeV},\qquad f_\pi=0.130~{\rm GeV},\nonumber\\
D&=0.80,\qquad F=0.47,\qquad
|\alpha_H|=0.0112~{\rm GeV}^3.
\label{eq:protoninputs}
\end{align}
with $A_R=2.5$ and $V_{ud}=0.974$.  Replacing $M_X$ by $\rho_XM_G$ gives
\begin{equation}
\tau(p\to e^+\pi^0)=\tau_0\rho_X^4,
\qquad
\tau_0=1.84\times10^{36}~{\rm yr},
\label{eq:protonscaling}
\end{equation}
for the unified coupling in \Eq{eq:gaugeresult}.  The limit in \Eq{eq:mxbound} follows directly from \Eq{eq:protonscaling}.

For dimension-five decay, the physically relevant quantity before dressing is the propagator element $\mathcal P_T$ defined in \Eq{eq:c5effective}.  If a future Higgs-sector completion yields, for example, a pseudo-Dirac sub-block
\begin{equation}
M_T=\begin{pmatrix}0&M\\M&\mu_R\end{pmatrix},
\label{eq:pseudodiracconditional}
\end{equation}
then $(M_T^{-1})_{11}=-\mu_R/M^2$.  This algebra illustrates how an approximate triplet-number symmetry can suppress the coefficient, and \Eq{eq:pseudodiracconditional} serves only as an algebraic illustration of the propagator suppression.  A quantitative $p\to K^+\bar\nu$ prediction requires deriving $M_T$ from the full $SU(6)$ Higgs superpotential and rotating the associated Yukawa matrices into the fermion mass basis.

\section{Explicit selection-rule}\label{app:operatorledger}
For each chiral monomial, the selection rules require $(R,P_\nu,q_G)=(2,0,0)$, with charges understood modulo $(4,2,3)$ respectively. The K\"ahler terms instead require $(0,0,0)$, with opposite $R$ and $Z_3^G$ charges for conjugate superfields. The relevant results are
\begin{center}
\begin{tabular}{@{}lccc@{}}\toprule
operator & $R$ & $P_\nu$ & allowed\\\midrule
$\bar6\Phi N_s$ & $2$ & $1$ & no\\
$\bar6'\Phi N_s$ & $2$ & $0$ & yes\\
$15\bar6'\bar H_6$ & $2$ & $1$ & no\\
$N_sN_s$ & $2$ & $0$ & yes\\
$\bar H_6\Phi$ & $2$ & $0$ & yes\\
$X\bar\Phi\Phi$ & $2$ & $0$ & yes\\
$S^3$ & $2$ & $0$ & yes\\
$SH_6\bar\Phi$ & $2$ & $0$ & yes\\
$P_0\bar6\bar6'H_{15}$ & $2$ & $0$ & yes\\
$P_2 15\bar6'\bar\Phi$ & $2$ & $0$ & yes\\\bottomrule
\end{tabular}
\end{center}
All entries in this table have zero $Z_3^G$ charge when the two illustrative $P$ spurions have $q_G(P_0)=q_G(P_2)=0$. Both $X^\dagger N_sN_s$ and $X^\dagger(\bar\Phi\Phi)N_sN_s$ have K\"ahler charges $(0,0,0)$ and are allowed by the original assignments.

\bibliography{references}
\end{document}